\documentclass[preprint,preprintnumbers,superscriptaddress,amsmath,amssymb]{revtex4-1}
\usepackage{graphicx,color,bm,natbib, placeins}

\usepackage{graphicx}
\usepackage{natbib}
\usepackage{color}
 \usepackage{amsbsy}
   \usepackage{amssymb}

\newcommand\Pen{\mbox{\textrm{Pe}}}  

\newsavebox{\astrutbox}
\sbox{\astrutbox}{\rule[-5pt]{0pt}{20pt}}

\newcommand{\bi}{\begin{itemize}}
\newcommand{\ei}{\end{itemize}}
\newcommand{\ben}{\begin{enumerate}}
\newcommand{\een}{\end{enumerate}}

\newcommand{\bfi}{\begin{figure}[hbtp]}
\newcommand{\efi}{\end{figure}}
\newcommand{\dr}{\partial}
\newcommand{\beq}{\begin{equation}}
\newcommand{\eeq}{\end{equation}}
\newcommand{\beqar}{\begin{eqnarray}}
\newcommand{\eeqar}{\end{eqnarray}}

 \newcommand{\vepsil}{\varepsilon}
 \newcommand{\vepsi}{\varepsilon}
 \newcommand{\ba}{\begin{array}}
\newcommand{\ea}{\end{array}}
\newcommand{\St}{{\rm St}}

\renewcommand{\vec}{\mathbf}

\begin{document} 

\hfill{\it J. Fluid Mech.\  \textit{\textbf{744}}, 183 (2014)}

\title{Inertial particle trapping in an open vortical flow} 

\author{Jean-R\'egis Angilella} 
\affiliation{\small Universit\'e de Caen et de Basse Normandie, LUSAC, Cherbourg, France }

\author{Rafael D. Vilela} 
\affiliation{\small $\mbox{Centro de Matem\'atica, Computa\c c\~ao e Cogni\c c\~ao, Universidade Federal do ABC (UFABC),}$ Santo Andr\'e-SP, 09210-170, Brazil}

\author{Adilson E. Motter} 
\affiliation{\small $\mbox{Department of Physics and Astronomy, Northwestern University, Evanston, IL 60208, USA}$}

\begin{abstract} \baselineskip 13.5pt
Recent numerical results on advection dynamics have shown that particles denser than the fluid can remain trapped indefinitely in a bounded region of an open fluid flow. Here,  we investigate this counterintuitive phenomenon both numerically and analytically to establish the conditions under which the underlying particle-trapping attractors can form. We focus on a two-dimensional open flow composed of a pair of vortices and its specular image,  which is a system we represent as a vortex pair plus a wall along the symmetry line. Considering particles that are much denser than the fluid, referred to as {\it heavy particles}, we show that two attractors form in the neighborhood of the vortex pair provided that the particle Stokes number is smaller than a critical value of order unity.  In the absence of the wall, the attractors are fixed points in the frame rotating with the vortex pair, and  the boundaries of their basins of attraction are smooth. When the wall is present, the point attractors describe counter-rotating ellipses in this frame, with a period equal to half the period of one isolated vortex pair.  The basin boundaries remain smooth if the distance  from the vortex pair  to the wall is large. However,  these boundaries are shown to become fractal  if the distance to the wall is smaller than a critical distance that scales with the inverse square root of the Stokes number. This transformation is related to the breakdown of a separatrix that gives rise to a heteroclinic tangle close to the vortices, which  we describe using  a Melnikov function. For an even smaller distance to the wall, we demonstrate that a second separatrix  breaks down and a new heteroclinic tangle forms farther away from the vortices, at the boundary between the open and closed streamlines.  Particles  released in the {\it open} part of the flow can approach the attractors and be trapped permanently provided that they cross the two separatrices, which can occur under the  effect of  flow unsteadiness. Furthermore, the  trapping of heavy particles from the open flow is shown to be robust to the presence of  viscosity, noise, and gravity. Navier-Stokes simulations  for large flow Reynolds numbers show that  viscosity does not destroy the attracting points until vortex merging takes place, while simulation of  thermal noise shows that particle trapping persists for extended periods provided that the P\'eclet number is large. The presence of a gravitational field does not alter the permanent trapping by the attracting points if the settling velocities are not too large. For larger settling velocities, however, gravity can also give rise to a limit-cycle attractor next to the external separatrix and to a new form of trapping from the open flow that is not mediated by a heteroclinic tangle.
\end{abstract}


\maketitle \baselineskip 14.5pt

\section{Introduction}
\label{secIntro}

The motion of particles transported by a fluid flow can be very complex even when the particles are passive, the dynamics is non-brownian, and the flow is  laminar \citep{Arnold1965, Gautero1985, Maxey1986,Maxey1987pof,McLaughlin1988, Antonsen1991, Cartwright2010, Babiano2000,Falkovich2001}. Contributing to this complexity,  the trajectories of particles with small but finite inertia often deviate significantly from 
fluid-point trajectories. The prediction of particle evolution is therefore a challenging task in particle-laden flows. In the paradigmatic case of spherical particles with small  Reynolds numbers,  the Maxey-Riley equation \citep{Maxey1983,Gatignol1983} 
can be used to describe the particle dynamics provided that the fluid velocity field is known. For non-interacting particles, as considered in this study, the complexity of this dynamics is mainly due to the spatial and temporal dependencies of the fluid velocity, which are strongly nonlinear in general.  

Previous theoretical analyses reported evidence of particle accumulation in well-defined regions of both laminar flows
 \citep{Maxey1987pof,Rubin1995,Vilela2007,Sapsis2010,Pushkin2011} and turbulent or random flows  \citep{Squire02,Fessler1994,Balkovsky2001,Bec2003,Mehlig2005,Duncan2005,Wilkinson2007,Wilkinson2010,Fouxon2012}. Particle clustering can occur even when the fluid itself is incompressible, and this is a property of major importance for the understanding  of many natural and industrial advection processes \citep{Barge1995,Cuzzi2001,Falkovich2002,Pasquero2003,Liu2010,Meyera2011}.   In bounded or periodic domains, this clustering behavior may be expected since particles have dissipative dynamics due to their inertia and dissipation can give rise to attractors. In  {
 {\it closed}
 } vortical flows, such attractors 
 tend to be associated with inward motion in the case particles less dense than the fluid---so-called bubbles---and with outward (but necessarily bounded) motion in the case of particles denser than the fluid  \citep{Maxey1987pof}---also known as aerosols.  Similar phenomenology is expected, and actually observed, for bubbles in {\em open} vortical flows \citep{Benczik2002,Benczik2003}. 

For aerosols, however,  the possibility of permanent clustering of particles in open flows is far less clear. This is  
the case
not only because particle motion is no longer constrained to be bounded but also because the same fluid velocity fields that have the potential to generate attractors tend to centrifuge denser particles away. Nevertheless, such attractor formation and consequent particle clustering has been shown to be possible for aerosols due to ``interactions" between coexisting vortices  \citep{Vilela2007}. This was demonstrated, for example, in numerical simulations of the open flow defined by leapfrogging vortices, where aerosols are trapped permanently by attracting points 
in the neighborhood of the vortices. The goal of the present paper is to determine both numerically {\it and} analytically the 
fluid and particle
conditions under which such attracting  
sets
exist, and investigate their properties as well as the properties of the associated basins of attraction. 

\begin{figure} 
        \centering
               \includegraphics[width=0.9\textwidth]{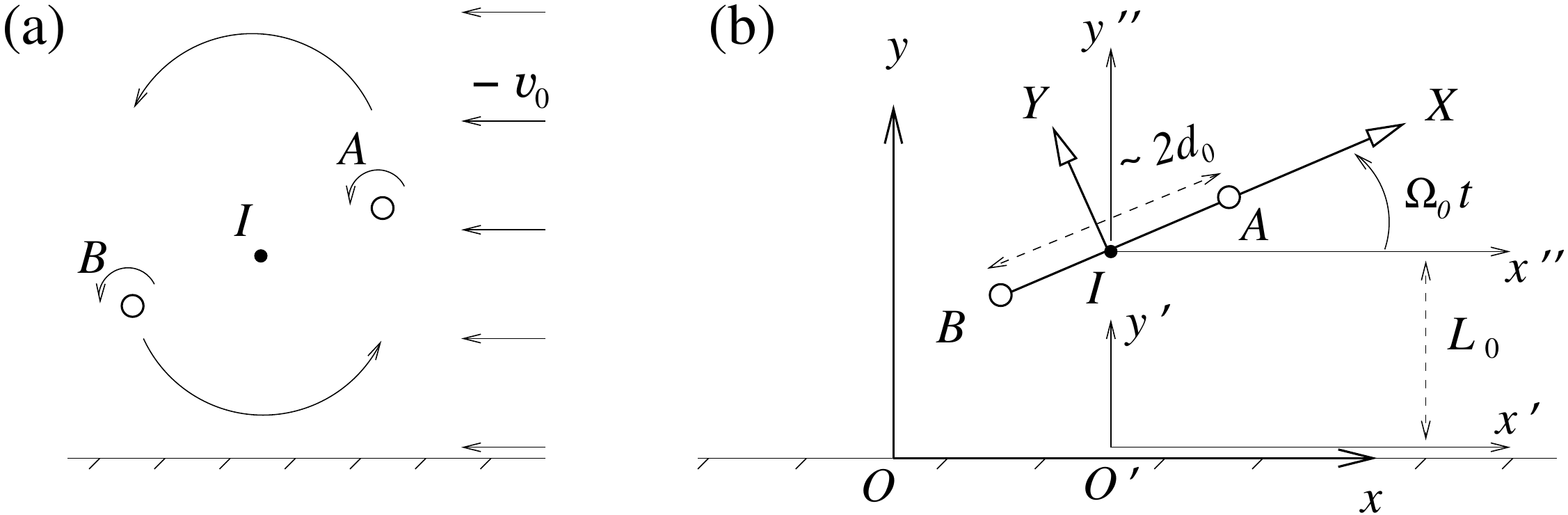}  
\caption{\baselineskip 13pt
Sketch of the vortex pair considered in this study.   (a) Symbols $A$ and $B$ denote co-rotating point vortices of same 
strength $\Gamma$.  To first order, the center of vorticity $I$ moves to the right with 
constant velocity $v_{{}_0}$  equal to $2\varepsilon\Omega_0 d_0 \equiv \Gamma/(2\pi L_0)$. (b) The vortices are separated from each 
other by an average distance $2d_0$,  
and  $I$  is at a distance $L_0$ from a wall (which can be interpreted
as a symmetry line) represented by the $Ox$ axis.  Here, $xOy$ is a coordinate system of the laboratory frame,  $x'O'y'$ and $x''Iy''$ are coordinate systems of the non-rotating frame translating with velocity $v_{{}_0}$, and $XIY$ is a coordinate system of the rotating frame.}
        \label{coords}
\end{figure}

Here, we consider small spherical particles much denser than the fluid, referred to as {\it heavy particles}, which capture the essential features of the problem while making it amenable to mathematical treatment.  We  focus on a system formed by two point-vortex pairs separated by a symmetry line, which is an open flow system that we represent as a single vortex pair plus a wall at the symmetry line (Fig.\ \ref{coords}(a)). 
Accordingly, the flow Reynolds number is assumed to be much larger than one (inviscid fluid approximation), even though the particle Reynolds number, based on the slip velocity and on the particle diameter, will be assumed to be small throughout this paper.
 We concentrate on the limit  
\begin{equation}
\vepsil = \frac{d_0}{L_0} \ll 1,
\end{equation}
where $d_0$ is the average half-distance between the vortices  and $L_0$ is the distance from the center of vorticity of the vortex pair to the wall (Fig.\ \ref{coords}(b)).  Due to the presence of the wall, the center of vorticity moves with respect to the distant fluid with velocity $v= \Gamma/(2\pi L_0)$ to first order in $\vepsil$, where $\Gamma$ is the strength of each vortex. In the small-$\vepsil$  regime, we can generalize analytical results on particle accumulation previously established for the closed-flow system defined by isolated (non-translating) point-vortex pairs with identical strengths \citep{Angilella2010}. Our analysis is partially based on using a perturbative fluid velocity field with respect to  $\vepsil$, where the case of an isolated vortex pair corresponds to $\vepsil =0$ (i.e., the absence of the wall) in our system.

In the reference frame translating with the center of vorticity, this flow exhibits open streamlines separated from closed streamlines by a separatrix formed by the invariant manifolds of two stagnation points (Fig.\ \ref{ComparePsiAsym4VortEpsi0.25}).  A central part of this work concerns the demonstration that, under appropriate conditions, heavy particles from the open flow can approach the vortices and be captured by attracting points in their neighborhood. In addition to this external separatrix, which we denote $\Sigma_3$, we anticipate that there are three other internal separatrices, denoted   $\Sigma_i$ for $i=0,1,2$,  which are located in the very neighborhood of the vortices and will be analyzed in the reference frame rotating with the vortex pair.  
These separatrices too will be shown to play a key role in the dynamics of inertial particles.
 
The motion of particles in this open vortical flow is investigated in Sec.\ \ref{secAsym},  where we show that attracting points exist even if $\vepsil >0$ provided that the particle Stokes number is small. In Sec.\ \ref{4vortex}, we show that the boundaries of the corresponding basins of attraction, which are smooth for $\vepsil =0$, become fractal if $\vepsil$ is above a critical value that decreases with increasing Stokes number. The occurrence of trapping from the open flow---for particles released far ahead of the vortex system in the upstream flow---is established and  analyzed in Sec.\ \ref{sepS1S2}.  In Sec.\ \ref{simulns}, 
we show that particle trapping is largely robust to the effects of gravity, viscosity, and noise.
In the same section we also show that gravity can induce the formation of a new (limit-cycle) attractor, that potential flow theory provides a good approximation to predict heavy particle dynamics preceding vortex coalescence,  and that noise can often enhance (rather than suppress) particle trapping.
 Final remarks are presented in Sec.\ \ref{concl}. We use no-slip initial conditions in all simulations (i.e., the particles are released with velocity equal to the local fluid velocity), which corresponds to 2-dimensional slices of the basins of attraction and nevertheless reveals  
geometric
properties of the full basins.

\section{Trapping of heavy particles near vortices}
\label{secAsym}

For heavy particles, as considered here, it has recently been shown via analytical calculations that a system comprised of two co-rotating identical point vortices has two fixed-point attractors in the rotating frame for Stokes numbers smaller than $2-\sqrt{3}$.  This holds true when the vortex pair is isolated, forming a closed fluid flow system since in this case the center of vorticity does not translate with respect to the fluid  \citep{Angilella2010}. The presence of a wall, on the other hand, allows the fluid to translate with respect to the vortex pair. This leads to a fundamentally different physical situation, in which the fluid flow system can now be open. In this section, by focusing on the velocity field in the neighborhood of the vortex pair, we study the persistency of the attractors and the properties of their attraction basins as a function of the Stokes number and distance of the vortices from the wall.  In particular, we establish a relation between the emergence of fractal basin boundaries and the breakdown of a separatrix in the neighborhood of the vortices.

\subsection{Perturbative internal fluid velocity field}

\begin{figure} 
        \centering
       \includegraphics[width=0.6\textwidth]{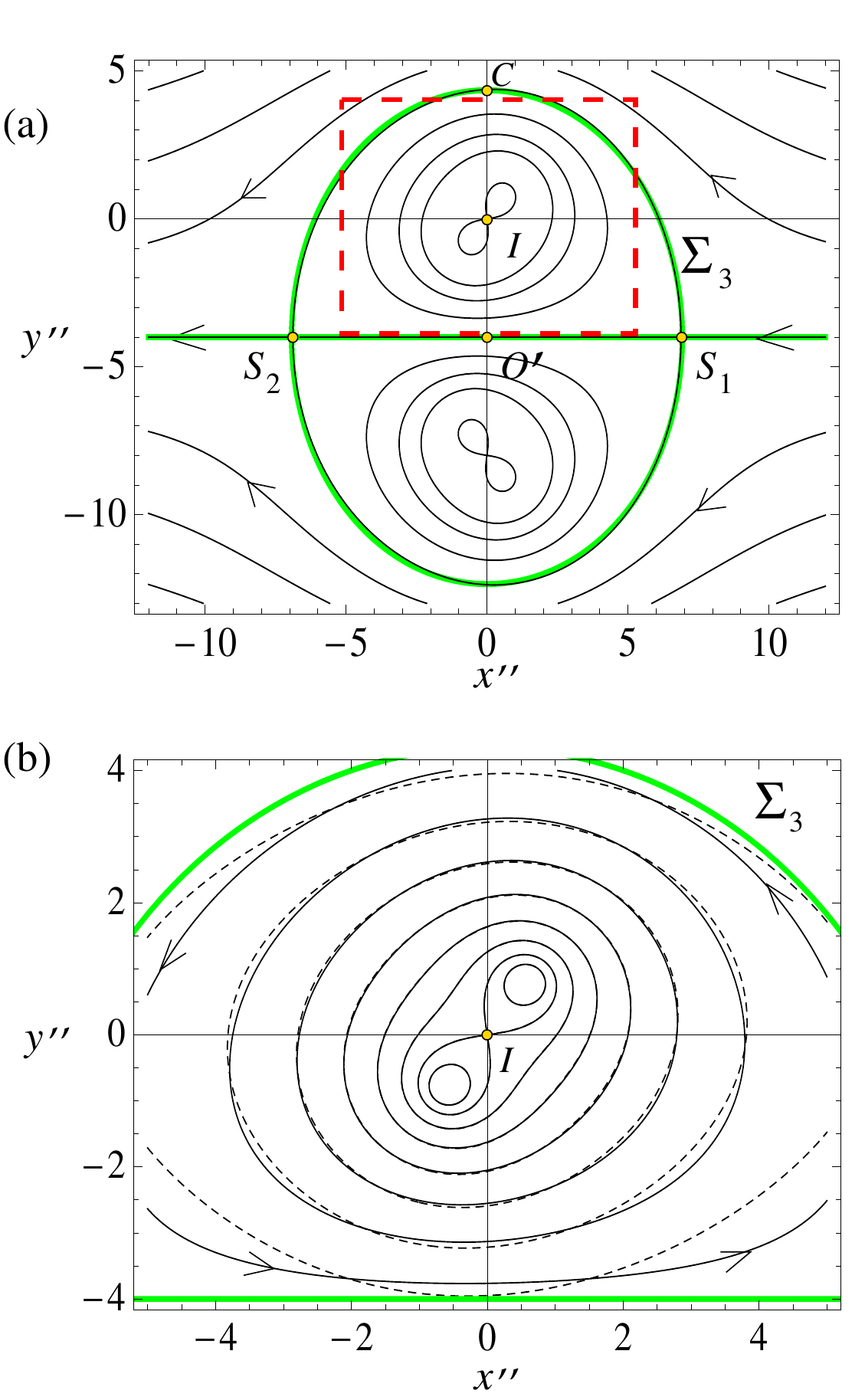}  
  \caption{\baselineskip 13pt
  Streamlines of the vortex system.
(a) Streamlines of the exact 
four-vortex
potential flow  in the frame translating with the vortices for  $\vepsil = 0.25$.
The bold curve (green) corresponds to the separatrix $\Sigma_3$ (and its mirror image) %
 between closed and open streamlines, which is associated with the fixed points $S_1$ and $S_2$.
(b) Magnification of the dashed rectangle of panel (a), showing the internal perturbative flow described by Eq.\ (\ref{psiapprox}) (dashed lines) on top of the  exact potential flow (solid lines). In both panels, the streamfunction isolines are taken at a particular time and are equispaced.
} 
\label{ComparePsiAsym4VortEpsi0.25}
\end{figure}

We first recall results of  \cite{Angilella2011}, where a perturbative expansion in $\vepsil$ was used to calculate the velocity field of the fluid for small $\vepsil$, when the vortex pair $(A,B)$ is distant from the wall. For $\vepsil = 0$ (i.e., in the absence of the wall), the vortices rotate around their center point $I$ with an angular velocity $\Omega_0 =  {\Gamma}/{(4 \pi d_0^2)}$, where the distance $2d_0 = |AB|$ between the vortices remains constant over time.  We make use of $\Omega_0$ and $d_0$ to set our equations non-dimensional in this section. The non-dimensional vortex strength is therefore equal to $4 \pi$.  For $\vepsil > 0$ (i.e., in the presence of the wall), the streamfunction is the sum of the flow induced by the two vortices plus the flow induced by the two mirror vortices, as illustrated in Fig.\ \ref{ComparePsiAsym4VortEpsi0.25}(a).  Under the effect of the mirror vortices, the point $I$ will translate in the $x$-direction  with a non-dimensional velocity equal to $2 \vepsil + O(\vepsil^3)$.  In the neighborhood of $(A,B)$, the contribution from the mirror vortices is a perturbation taking the form of a straining flow.  The resulting non-dimensional streamfunction  in the reference frame $x''Iy''$ translating with $I$ at 
velocity $2 \vepsil$, reads
\beq
\psi_I(x'',y'',t) = \sum_{i=1}^2 - \mbox{ln}\left[ (x''-x''_i)^2 + (y''-y''_i)^2 \right] + \frac{\vepsil^2}{2}  ({x''}^{2}  -  {y''}^2 ) + O(\vepsil^3),
\label{psiapprox}
\eeq
where $(x''_i(t),y''_i(t))$ are the Cartesian coordinates of the vortices $(A,B)$. Because this streamfunction is valid near the vortices only, we refer to Eq.\ (\ref{psiapprox}) as an {\it internal perturbative solution}. Figure \ref{ComparePsiAsym4VortEpsi0.25}(b) shows a comparison for $\vepsil = 0.25$ between the exact potential flow induced by the four vortices  and the perturbative solution. Even though $\vepsil$ is not very small, the streamlines are essentially undistinguishable in the neighborhood of the vortices. We have checked that the agreement is also satisfactory for the values of the velocity.   Significant discrepancies start to appear at distances of about $3$ non-dimensional units from $I$. In particular, the stagnation points $S_1$ and $S_2$ appearing on the symmetry line are not captured by the internal perturbative model, since they are points where the contribution of the two upper and two lower vortices have equal amplitudes and opposite signs. 

One can verify that the dynamics of the vortices in this simplified flow satisfies $x''_1(t) = r(t) \cos \theta(t),\, y''_1(t) = r(t) \sin \theta(t)$, $x''_2 = -x''_1$, and $y''_2=-y''_1$, with  $r(t) = 1 +({\vepsil^2}/{2}) \cos 2 t$ and $\theta(t) = t - \vepsil^2 \sin 2 t$, plus terms of order $\vepsil^4$.  The distance $2 r(t)$  between the two vortices therefore oscillates with a period $\pi$ (half the period of the isolated vortex pair, where throughout this paper we define the period as the time for each vortex to return to its original position in the coordinate system $x''Iy''$).  In addition, the angular velocity  of the vortices around $I$ is affected by a perturbation  with period $\pi$. This periodic forcing corresponds to the effect of a wall-induced straining flow on the vortices (see also \cite{Carton2002} and \cite{Maze2004} for vortex pairs in a straining flow).

It is well known that in the absence of the wall the velocity field  is steady when observed in the  rotating frame of the two vortices. It is therefore useful to re-write the streamfunction (\ref{psiapprox}) in the coordinate system $XIY$ defined on this rotating frame: $\psi_r(X,Y,t) = \psi_I(x'',y'',t) + (X^2+Y^2)/2$.   Assuming that the axes  $IX$ and $IY$ correspond to $Ix''$ and $Iy''$ at $t=0$,  the perturbative streamfunction reads 
\beq
\psi_r(X,Y,t) =  \psi_{r0}(X,Y) + \varepsilon^2  \psi_{r2}(X,Y,t),
\label{psir}
\eeq
where $Z=X+iY$, $\psi_{r0}(X,Y) = -2 \mbox{ ln}|Z^2-1| + |Z|^2/2$, and $\psi_{r2}(X,Y,t)=-\frac{2}{|Z^2-1|^2}[(Y^2-X^2+1)\cos2t +4XY\sin2t]+\frac{1}{2}[(X^2-Y^2)\cos2t -2XY\sin2t]$  \citep{Angilella2011}. The fluid velocity field in the rotating frame then takes the form 
\beq
\label{innerVf}
\vec W_f(X,Y,t) = \frac{\dr \psi_r}{\dr Y} \hat{\vec X} - \frac{\dr \psi_r}{\dr X} \hat{\vec Y} =\vec W_{0}(X,Y) + \varepsilon^2 \vec W_{2}(X,Y,t),
\eeq
where $ \vec W_{2}(X,Y,t) =   \vec W_{2c}(X,Y) \cos 2 t + \vec W_{2s}(X,Y) \sin 2 t$,
and the expressions of the steady fields $\vec W_{2c}$ and $\vec W_{2s}$ are obtained by differentiating with respect to the spatial variables
 the coefficients of $\cos 2 t$ and $\sin 2 t$ appearing in $\psi_{r2}$. In the next subsection, we investigate for the first time the motion of heavy particles in this flow.

\subsection{Particle motion and attracting points}
\label{partmotion}

In the rotating reference frame, the equation of motion
for such a  heavy particle is \citep{Maxey1983,Gatignol1983}
  \beq
\frac{d^2 \vec{X}_p}{dt^2} = \frac{1}{\St}\left( \vec W_f - \frac{d \vec{X}_p}{dt} \right) + \vec X_p - 2 \hat{\vec z} \times \frac{d \vec{X}_p}{dt},
\label{PtclMotion}
\eeq
where $\vec X_p$ is the position vector of the particle, $\St = \Omega_0 \tau_p$ is the Stokes number,  $\tau_p$ is the particle 
relaxation time,
 and $\hat{\vec z}$ is the unit vector along the $z$-axis (perpendicular to the plane).  The first term on the right side of this equation is the drag force,  the second term is the centrifugal force, and the last term is the  Coriolis force. In the rotating frame, the  force  due to the undisturbed flow also contains terms equal to the opposite of the Coriolis and centrifugal forces acting on the fluid. These forces, as well as the added mass,  history, buoyancy, and lift forces \citep{Michaelides1997},  have not been taken into account since they are negligible for sufficiently small and heavy particles. 
The settling velocities are assumed to be negligible throughout the paper, except in Sec.\ \ref{secgravity}, where the effect of gravity is considered in detail.

When $\vepsil=0$,  there are four equilibrium positions (in addition to $I$) where the particle drag balances the centrifugal force if 
$\St < 2-\sqrt 3$ or $\St > 2+\sqrt 3$. Two of them are stable if $\St < 2-\sqrt 3$, while the others are always unstable.
The stable points, which we denote $\pm \vec X_{eq}$, are symmetric with respect to $I$; their polar 
coordinates, defined by $\pm\vec X_{eq} \cdot \hat{\vec X} = R \cos \Theta$ and $\pm\vec X_{eq} \cdot \hat{\vec Y} = R \sin \Theta$,  read
\begin{eqnarray}
R &=& \sqrt{\cos 2 \Theta + \frac{\sin 2 \Theta}{ {\rm St}}},\\
 \Theta &=& \pm \frac{\pi}{2}\mp\frac{1}{2} \mbox{arcsin} \frac{4 {\rm St}}{ 1  +  {\rm St}^2 }.
\end{eqnarray}

These equilibrium points no longer exist when $\vepsil > 0$, since the flow is no longer time-independent in the rotating frame $XIY$. Nevertheless, particles can be attracted to {\it moving} stable points in the vicinity of the equilibrium points $\pm \vec X_{eq}$ that exist for $\vepsil=0$ (they are in fact limit cycles in the extended phase space that includes time as one of the dimensions). To analyze this effect, we employ the method used by \cite{IJzermans2006} for particles in a periodic box. We focus on the vicinity of $\vec X_{eq}$, as the corresponding considerations for $-\vec X_{eq}$ follow immediately by symmetry. Replacing ${\vec X}_p(t) =  \vec X_{eq} + \vec h(t)$ in the equation of motion (\ref{PtclMotion})  and performing a Taylor expansion with respect to $\vec h(t)$,  we obtain
\beq
\frac{d^2 \vec h}{dt^2} = \frac{1}{{\rm St}} \left[ 
{
\vec h \cdot \nabla \vec W_{0,eq} 
}
 + \vepsil^2 \left(\vec W_{2c} (\vec X_{eq}) \cos 2 t + \vec W_{2s} (\vec X_{eq}) \sin 2 t  \right) - \frac{d  \vec h}{dt} \right]  + \vec h - 2 \hat{\vec z} \times \frac{d  \vec h}{dt} ,
\label{eqh}
\eeq
where $\nabla \vec W_{0,eq}$ is the gradient tensor  
of the fluid velocity $\vec W_0$ at $\vec X_{eq}$. In this expression, terms of order $|\vec h|^2$ and  $\vepsil^2 |\vec h|$ have been neglected. The solution of this linear non-homogeneous equation is the sum of a particular solution $\vec h_a(t)$ of the full equation and  the general solution $\vec h_b(t)$  of the homogeneous part of the equation.

By setting $\vec h_a(t) = \vec p \cos 2 t + \vec q \sin 2 t$ in Eq.\ (\ref{eqh}), we are led to the following conditions for the coefficients of $\cos 2 t$ and $\sin 2 t$ \citep{IJzermans2006}: 
\beq
 {\mathbf L}   \vec p - {\mathbf M}   \vec q = -\frac{\vepsil^2}{\St}\vec W_{2c} (\vec X_{eq}) ,
\eeq  
\beq
{\mathbf L}  \vec q + {\mathbf M}   \vec p =   -\frac{\vepsil^2}{\St}\vec W_{2s} (\vec X_{eq}) ,
\eeq  
with
\beq
{\mathbf L} = \frac{1}{{\rm St}}   \nabla \vec W_{0,eq}   + 5 {\mathbf I} 
\eeq
and
\beq
{\mathbf M} = \frac{2}{{\rm St}}   {\mathbf I}  + 4 {\mathbf A} ,
\eeq
where
$\displaystyle{
{\mathbf A} = \left(\ba{cc}
0 & -1\\
 1 & ~~0\\
\ea\right)
}$ and  ${\mathbf I}$ is the identity matrix.
One can easily check that ${\mathbf M}$ and  ${\mathbf L}$ are  invertible, where the latter follows from  the eigenvalues of the attracting points having strictly nonzero real parts (and hence  $\nabla \vec W_{0,eq} $ being non-singular).  It follows that
\beq
\left( {\mathbf M}^{-1} {\mathbf L} +  {\mathbf L}^{-1} {\mathbf M} \right)  \vec q   =  \frac{\vepsil^2}{\St} \left( {\mathbf L}^{-1} \vec W_{2c} (\vec X_{eq}) - {\mathbf M}^{-1}\vec W_{2s} (\vec X_{eq})\right) ,
\eeq  
\beq
\left( {\mathbf M}^{-1} {\mathbf L} +  {\mathbf L}^{-1} {\mathbf M} \right) \vec p   =  -\frac{\vepsil^2}{\St} \left( {\mathbf L}^{-1} \vec W_{2s} (\vec X_{eq}) + {\mathbf M}^{-1}\vec W_{2c} (\vec X_{eq})\right).
\eeq  
By solving this system, one can obtain a closed form for $\vec p$ and $\vec q$, and this provides a particular solution to Eq.\ (\ref{eqh}). An approximate expression can be found  for  ${\rm St} \ll 1$ by performing a Taylor expansion of both $\vec X_{eq}$ and the various matrices in terms of ${\rm St}$. To first order in ${\rm St}$, the solution reads
\beq
\vec h_a(t) = \sqrt 3 \vepsil^2 \left(\frac{2124}{169} {\rm St} \cos 2 t - \frac{15}{13} \sin 2 t\right) \hat{\vec X} - \sqrt 3 \vepsil^2 \left(\frac{27}{26}  \cos 2 t + \frac{752}{169} {\rm St}\sin 2 t\right) \hat{\vec Y}. 
\label{ellipse}
\eeq
This solution corresponds to particles on an elliptic trajectory rotating with  period $\pi$ in the clockwise direction  around $\vec X_{eq}$  in the rotating frame. 

The general solution $\vec h_b(t)$ of the homogeneous part of   
 Eq.\ (\ref{eqh})  is nothing more than  
 a perturbation around $\vec X_{eq}$ for $\vepsil=0$. Because the corresponding eigenvalues  have strictly negative real parts for $\St < 2-\sqrt 3$  \citep{Angilella2010},  we infer that  $\vec h_b(t) \to 0$ as $t\rightarrow\infty$. We thus conclude that particles are attracted to the vicinity of the points
$\pm\vec X_{eq} + \vec h_a(t)$, where $\vec h_a(t)$ is the particular periodic solution approximated by Eq.\ (\ref{ellipse}). Note that, for capturing the effect of the wall, the solution $\vec h_a(t)$ cannot be anticipated from the existing literature on isolated vortex pairs.

\begin{figure}
        \centering
                \includegraphics[width=0.65\textwidth]{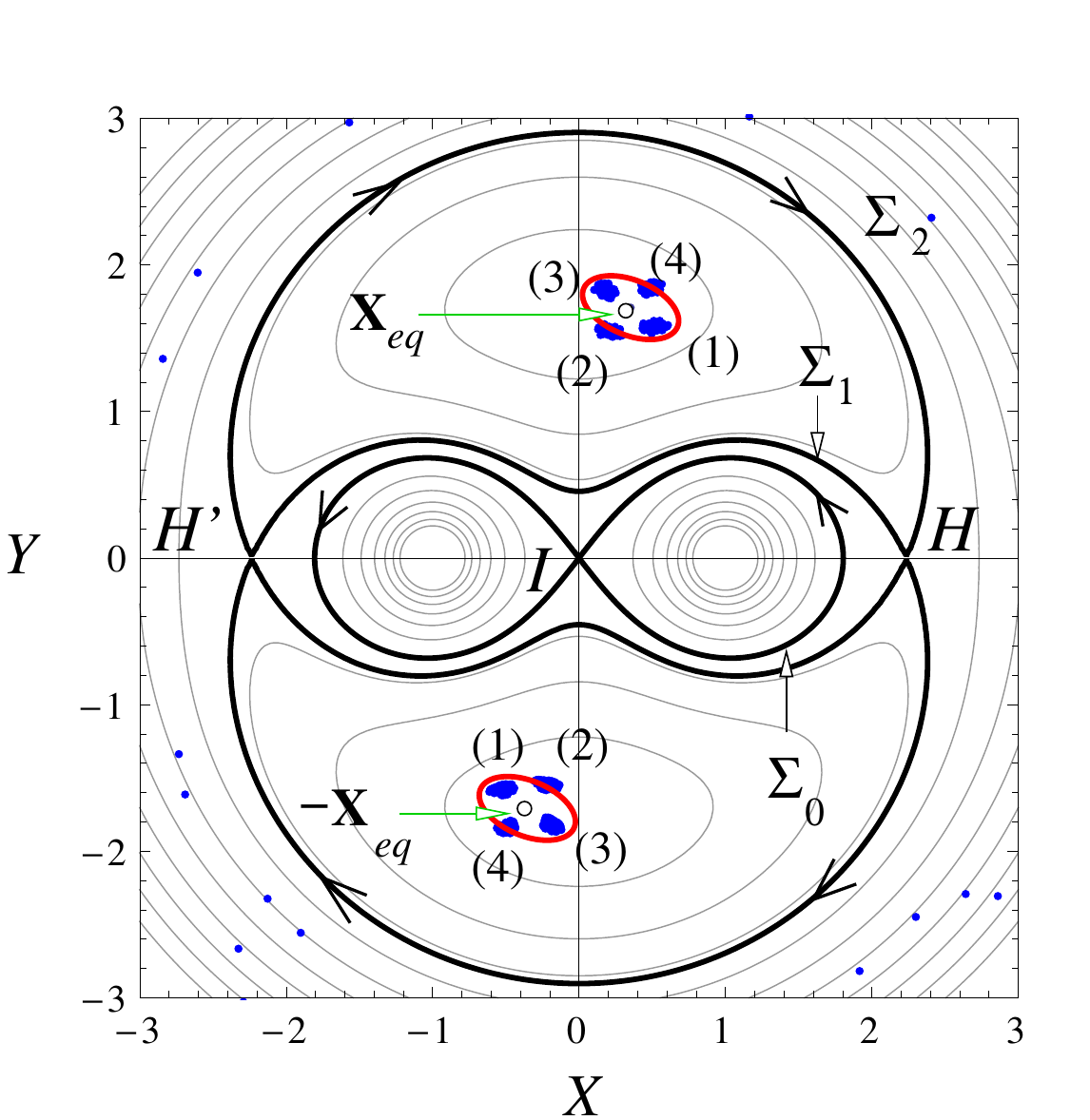}  
 \caption{\baselineskip 13pt
 Periodic attractors in the rotating reference frame.
The blue dots correspond to the simulated final positions of heavy particles with ${\rm St} = 0.1$ transported by the flow defined by $\vepsil = 1/3$. The red ellipses (solid lines) correspond to the analytical prediction in Eq.\ (\ref{ellipse}) for attractors projected on the physical space. 
The simulations assume that the particles have initial velocity equal to the fluid velocity and are initially uniformly distributed in the region shown. 
The numbers  (i) next to the particle clusters indicate the time $t_f-(\frac{4-i}{4})\pi$   ($i=1, \dots  4$) at which the particles are  observed, where the final time $t_f=14\times 2\pi$ corresponds to $14$ turnover times of the vortex pair. 
 Particles away from the attracting points correspond to initial conditions outside the basins of attraction.  Regular lines represent equispaced streamlines, bold lines represent the separatrices $\Sigma_0$, $\Sigma_1$, and $\Sigma_2$, and open circles indicate the stable equilibrium points $\pm {\vec X}_{eq}$ in the limit of vanishing $\vepsil$ (i.e., in the absence of the wall). }
\label{nuageLeapFIG}
\end{figure}

Figure \ref{nuageLeapFIG} shows a simulation of a particle cloud for $\vepsil = 0.33$ and ${\rm St} = 0.1$. Initially, the particles are distributed uniformly in the square $[-3,3]\times[-3,3]$,
which includes the vortex pair,
and have velocity equal to the local fluid velocity.  
The particle cloud is shown after $14$ periods at four different instants (blue dots). 
We indeed observe that particles are attracted by two moving points rotating clockwise around $\pm\vec {X}_{eq}$ with a trajectory that is close to the elliptic  orbit predicted theoretically in Eq.\ (\ref{ellipse}). The parameter $\vepsil$ has been taken rather large here to facilitate visualization, and agreement with the theoretical predictions only improves for smaller $\vepsil$.

\subsection{Crossing of the internal separatrix}
\label{MelnikSt}

We now turn to the emergence of chaos in the particle dynamics due to the perturbation of homoclinic  and/or heteroclinic orbits.  An homoclinic orbit in which branches of the stable and unstable manifolds of a fixed point (or, more generally, of a periodic orbit) coincide is a common structure in nonchaotic systems; the heteroclinic counterpart corresponds to the situation in which a branch of the stable manifold of one orbit coincides with a branch of  the unstable manifold of another orbit. Generic perturbations of such systems typically lead such branches to no longer coincide. If the manifolds associated with the perturbed homoclinic (heteroclinic) orbit(s)  
are found to intersect transversely at one point, then they will intersect transversely at infinitely many points, forming a homoclinic (heteroclinic) tangle and, in particular, giving rise to a chaotic set around the original manifold. These transverse intersections can be detected using the Melnikov method, where such intersections correspond to isolated odd zeros of an integral function---the Melnikov function---which provides a measure 
of the signed distance between the stable and unstable manifolds  \citep{Guckenheimer1983,Ottino1989}.

In order to proceed with our analysis of the particle dynamics in the internal perturbative flow, we first compare the order of magnitude of the various forces appearing in the equation of motion (\ref{PtclMotion}) with the approximate velocity field (\ref{innerVf}).   
Equation (\ref{PtclMotion}) contains two small parameters, namely
$\mbox{St}$ (accounting for inertia effects) and $\varepsilon$ (accounting for the wall effect).  
Clearly, if $\St \ll \varepsilon^2$,   the  velocity of the particle is only slightly modified by inertia. In contrast, if $\St \gg \varepsilon^2$, particles do not feel the effect of the wall,  as the corresponding oscillation of the vortices is weak. Therefore, we assume throughout that $\St =O(\varepsilon^2)$ and set 
\begin{equation}
\St = k\, \varepsilon^2,
\end{equation}
where $k$ is a constant of order one.
This condition is necessary to keep both the effect of the wall and the effect of  
inertia significant.
Then, taking a perturbative solution of the equation of motion in the form \citep{Maxey1987jfm,Haller2008,Haller2010}
\beq
\frac{d \vec{X}_p}{dt} = \vec W_{f}(\vec X_p,t)  + \St \left[ \vec X_p - 2 \hat{\vec z} \times  \vec W_{f}(\vec X_p,t) - \vec W_{f}(\vec X_p,t) \cdot \nabla \vec W_{f}(\vec X_p,t) \right]  + O(\St^2), 
\eeq
we obtain 
\beq
\frac{d \vec{X}_p}{dt} = \vec W_{0}(\vec X_p)  + \varepsilon^2 \vec W_{2}(\vec X_p,t)
+k \varepsilon^2 \left[ \vec X_p - 2 \hat{\vec z} \times  \vec W_{0}(\vec X_p) - \vec W_{0}(\vec X_p) \cdot \nabla \vec W_{0}(\vec X_p) 
\right] + O(\varepsilon^4).
\eeq
Under the given conditions, the dynamics of inertial particles is therefore equivalent to a Hamiltonian system perturbed by terms of order $ \varepsilon^2$. The unperturbed phase portrait is the same as the one of fluid particles, which is the well-known streamline diagram of co-rotating  point-vortex pairs shown in Fig.\ \ref{nuageLeapFIG}. As indicated in that figure, this flow  
has two heteroclinic orbits, $\Sigma_1$ and  $\Sigma_2$, forming separatrices  associated with the fixed points $H$ and $H'$ and has two homoclinic orbits, $\Sigma_0$, which form separatrices associated with the fixed point $I$.  The possibility of homoclinic and heteroclinic tangles in the internal perturbative flow can then be analyzed using the Melnikov method applied to these orbits.

Specifically, the Melnikov functions of the separatrices $\Sigma_i$ for $i=0,1,2$ will indicate whether, under the effect of the $\varepsilon^2$ perturbations, the invariant manifolds  associated with the various hyperbolic points will intersect transversely or not. Transverse intersections imply that a chaotic set exists  in the vicinity of $\Sigma_i$ and that particles may experience transient chaotic behavior in this region  
before converging to one of the attracting points or being centrifuged away (although this is not necessarily the case in general, our numerics do not indicate any other outcome for the flow and parameters considered here). 
To first order in $\varepsilon^2$, the transverse signed distance between the invariant manifolds associated with the separatrix $\Sigma_i$
at some  point $\vec X^* \in \Sigma_i$ is given by $d_i(t_0) = \varepsilon^2 a_i M_i(t_0)$, where $a_i$ is independent of $\varepsilon$ and $M_i$ is the Melnikov function.  For the separatrix $\Sigma_i$,  we obtain 
\begin{eqnarray}
M_i(t_0) = & \int_{-\infty}^{\infty}  {\dot  {\vec q}_i(t)}   \times \left[  \vec W_{2c}(\vec q_i(t)) \cos2(t+t_0) + 
{
  \vec  W_{2s}(\vec q_i(t))
  }
   \sin 2(t+t_0) \right] dt \nonumber\\
&+ k \int_{-\infty}^{\infty}   {\dot  {\vec q}_i(t)}  \times \left[ \vec q_i(t) -2 \hat{\vec z} \times  {\dot  {\vec q}_i(t)}  -  {\ddot  {\vec q}_i(t)} \right] dt ,
\label{Mit0}
\end{eqnarray}
where $t_0$ is the starting time of the stroboscopic map  $\vec X_p(t) \to \vec X_p(t+\pi)$,  
 and $\vec q_i(t)$ is a solution of the unperturbed system with $\vec q_i(0)=\vec X^*$. 
 As $t_0$ varies, the manifolds evolve and any intersection between them will be detected at $\vec X^*$.
  Because $\vec q_i(t)$ does not depend on $\vepsil$ or $\St$, we  compute this solution numerically for $i=0,1,2$ and use this solution to calculate the above integrals. Also, we make use of the fact that $\vec q_i(t)$ and its derivative are symmetric functions,
so that some of the integrals vanish. This leads to
\beq
M_i(t_0) = \alpha_i \sin 2 t_0 + k \, m_i ,
\label{main_melnikov}
\eeq
where   $\alpha_i$ and $m_i$ are purely numerical constants. The former constants  have been calculated in a previous work \citep{Angilella2011}: $\alpha_0 \approx -0.58$, $\alpha_1 \approx -0.89$, and $\alpha_2 \approx 7.3$. The latter constants are $m_0 \approx -42.1$,  $m_1 \approx -25.8$, and $m_2 \approx 8.3$.  Therefore, for any $\St$ or $\varepsilon$ larger than zero, the Melnikov function is no longer zero for each of the three separatrices, indicating that the stable and unstable manifolds no longer coincide. The pertinent question is then whether they intersect each other transversely.

The $\alpha_i$ constants reflect the influence of the time dependence of the fluid flow  on the splitting of the separatrices. The $m_i$'s account for the effect of the particles' inertia only.  Function $M_i(t_0)$ has no zeros if $k > \max_{i=0,1,2}  {|\alpha_i|}/{| m_i |} =  {|\alpha_2|}/{| m_2 |}$.  This is equivalent to the condition
\beq
\St> \St_{c_2} \equiv \frac{|\alpha_2|}{| m_2 |}\varepsilon^2,
\label{critere}
\eeq
where ${|\alpha_2|}/{| m_2 |} \approx 0.88$.
If this condition is fulfilled, the stable and unstable manifolds associated with the hyperbolic points $I$, $H$ and $H'$ of the stroboscopic map, which persist if $\varepsilon$ is small enough, split apart but do not intersect each other:  particles injected near the separatrices will evolve non-chaotically and eventually move either toward an attracting point or toward infinity.  
Conversely, if the inverse inequality is satisfied in Eq.\ (\ref{critere}), then it follows from Eq.\ (\ref{main_melnikov}) that $M_2(t_0)$ will have isolated odd zeros as a function of  $t_0$, and this implies the existence of a heteroclinic tangle. It is the separatrix $\Sigma_2$ that is represented in Eq.\ (\ref{critere}) because this separatrix is more sensitive than $\Sigma_0$ and $\Sigma_1$ to the presence of the wall: for a given Stokes number, if one increases $\varepsilon$ so that $\St<\St_{c_2}$,  the invariant manifolds of 
the separatrix $\Sigma_2$ will be the first to intersect transversely. If $\varepsilon$ is further increased, i.e., if the distance from the wall is further reduced, then  $\Sigma_{1}$ and $\Sigma_{0}$  will, in this order, give rise to chaotic sets by a similar mechanism.

\begin{figure}
        \centering
                \includegraphics[width=0.85\textwidth]{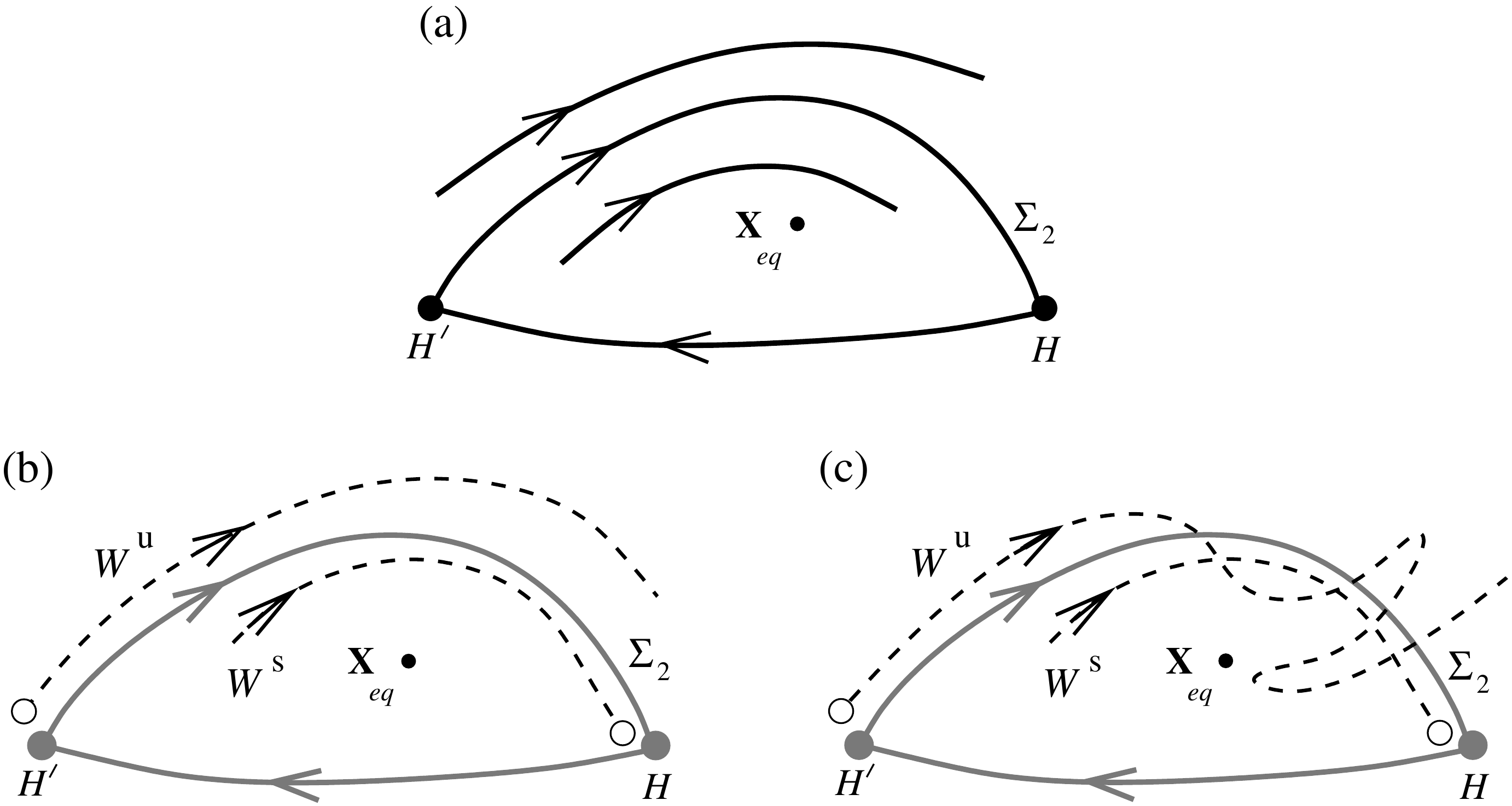}
 \caption{\baselineskip 13pt
 Sketch of impact of inertia and wall on the separatrix $\Sigma_2$.
 (a) The separatrix $\Sigma_2$ in the absence of both particle inertia and wall-induced flow unsteadiness. 
 In this case, the stable manifold $W^s$ of $H$ and unstable manifold $W^u$ of $H'$ 
 coincide with the separatrix.
 (b) Separation between the stable and unstable manifolds induced by particle inertia, when the inertia
 dominates over the wall effect. In this case, the manifolds do not intersect each
 other, but particles outside  $W^s$  cannot reach the attracting point $\vec X_{eq}$ because the
 particle-velocity field between $W^s$  and $W^u$  spirals outward.
 (c) Transverse intersections between the stable and unstable manifolds induced by the wall,
 when the wall dominates over the particle inertia. In this case, the particles from outside that are in the lobes
 bounded by $W^u$ can now cross $W^s$ and as a result they can in principle approach the attracting
 point.}
 \label{SketchManifolds}
\end{figure}

Figure \ref{SketchManifolds} summarizes these different behaviors. It shows the stable manifold $W^s$ and unstable manifold $W^u$ coinciding in the absence of both particle inertia and wall (Fig.\  \ref{SketchManifolds}(a)), split apart when particle inertia dominates (Fig.\  \ref{SketchManifolds}(b)), and intersecting each other transversely when the effect of the wall dominates (Fig.\  \ref{SketchManifolds}(c)). Very remarkably, particles outside the separatrix $\Sigma_2$ cannot reach the neighborhood of the attracting point $\vec X_{eq}$ when the invariant manifolds are split apart. This is the case because, as indicated in Fig.\  \ref{SketchManifolds}(b), the velocity field of particle dynamics points outward in the region between  $W^s$ and $W^u$. That is, the stable manifold is encircled by the unstable one, which corresponds to positive values for the Melnikov function $M_2$ (according to the convention for the signed distance adopted in this paper).  In this process,  $W^s$ can be regarded as a barrier to the transport of particles from the outside. However, it becomes possible for outside particles to reach the inner region (interior to $\Sigma_1\cup \Sigma_2$) when  the effect of the wall is dominant and induces transverse intersections between $W^s$ and $W^u$. Indeed, in this case, certain particles located outside the stable manifold---those in the lobes limited by $W^u$ \citep{RomKedar1990}---are transported to the other side of the stable manifold after one period of the stroboscopic map. This tangle, and hence the transient chaos that comes with it, is a necessary (albeit not sufficient) condition for outside particles to eventually approach the attracting point.

Finally, because the Melnikov functions $M_1$ and $M_0$ are negative for $\St/\varepsilon^2$ larger than $|\alpha_1|/| m_1 |$ and  $|\alpha_0|/| m_0 |$, respectively, the reciprocal argument applies to the separatrices $\Sigma_1$ and $\Sigma_0$. That is, the orientation of the particle-velocity field is such that these separatrices become permeable toward the interior of the  $\Sigma_1\cup \Sigma_2$ cycle as soon as their 
stable and unstable manifolds split apart  (in contrast with $\Sigma_2$, which requires the emergence of chaos, and hence larger $\varepsilon$,  to become permeable).
Physically, this occurs due to the centrifugation of the particles in the vicinity of the vortices.

These results can be interpreted also in terms of the distance between the vortices and the wall. Chaos exists in the vicinity of the vortices if and only if the vortex pair is placed below a critical distance  $L_c \sim d_0/\sqrt\St$ from the wall. Above this critical distance, the influence of the wall, and consequent oscillation in the inter-vortex separation, is too weak to induce chaos in the dynamics of  inertial particles.   On the other hand, since this critical distance scales as $\St^{-1/2}$, for any large but finite distance between the vortex pair and the wall, chaos will always manifest itself for sufficiently small heavy particles. As shown in the next section, the existence of a chaotic set critically impacts the geometry of the basin boundaries of the attracting points.  

\section{Fractal basin boundaries}
 \label{4vortex}
 
The appearance of a chaotic set in the vicinity of  separatrices  can drastically change the  boundaries of the basins of attraction, since inertial particles can move erratically in that region before either approaching an attracting point or being centrifuged away. This transiently chaotic dynamics imprints a signature in the geometry of the boundaries between the different basins of attraction.  
We thus expect that the boundaries of the attraction basins will be smooth when no chaotic sets are present around the separatrices but become fractal when such  sets exist and are sufficiently wide to be connected with the attraction basins.  

\begin{figure} 
        \centering
                 \includegraphics[width=0.8\textwidth]{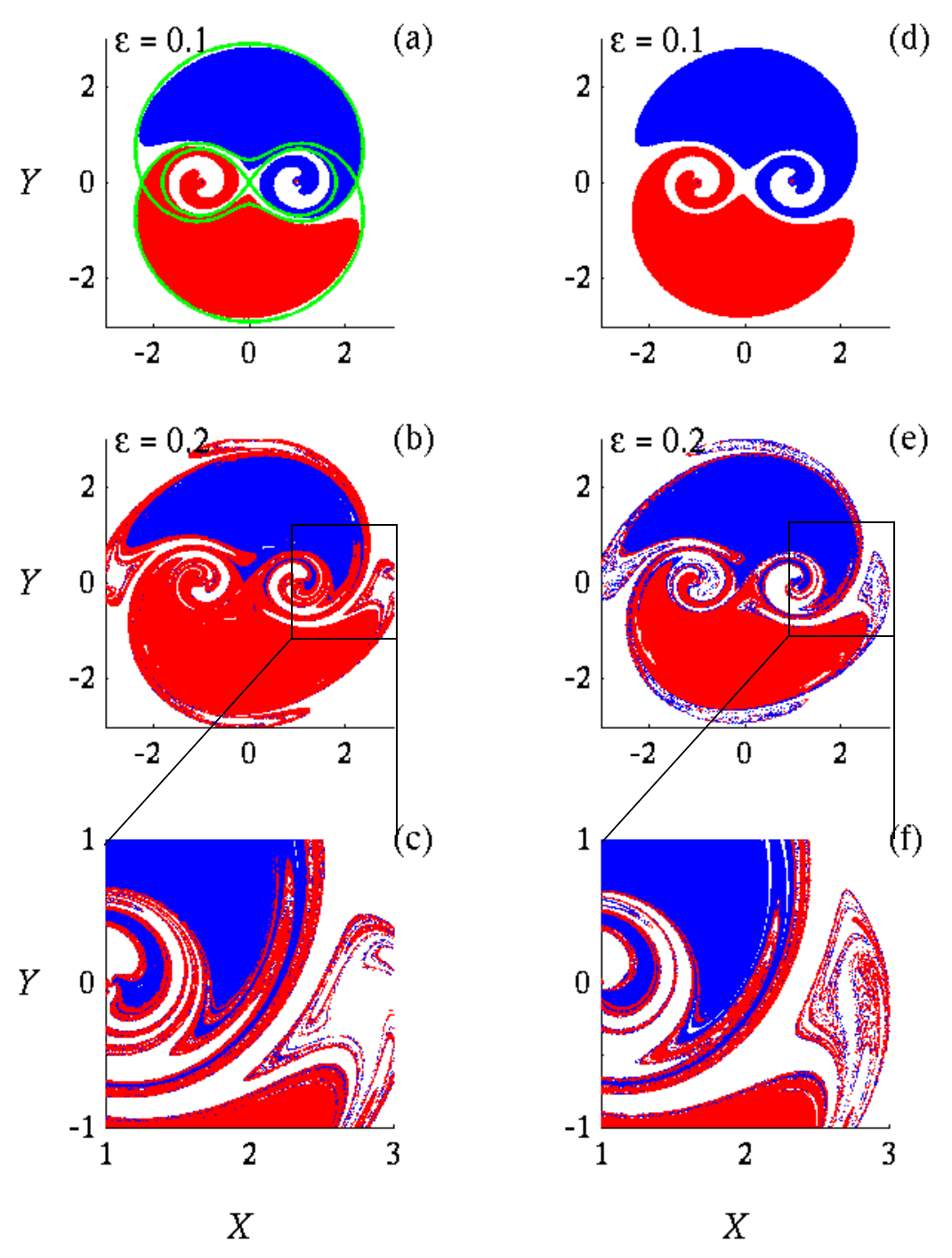} 
 \caption{\baselineskip 13pt
 Smooth versus fractal basin boundaries. Blue and red indicate the basins of attraction associated with the two periodic attractors for heavy particles for $\St = 0.02$ and initial velocity equal to the fluid velocity. (a-c)  Perturbative velocity field simulations. (d-f)  Exact potential flow simulations. The basin boundaries are smooth when the distance from the center of vorticity to the wall is large (a, d) and become fractal as this distance is reduced (b, e). Panels (c) and (f) are magnifications of the rectangles shown in panels (b) and (e), respectively. The continuous (green) lines in panel (a) correspond to the separatrices defined in Fig.\ \ref{nuageLeapFIG}.}
        \label{BasinsSt0.02}
\end{figure}

\begin{figure} 
        \centering
                \includegraphics[width=0.8\textwidth]{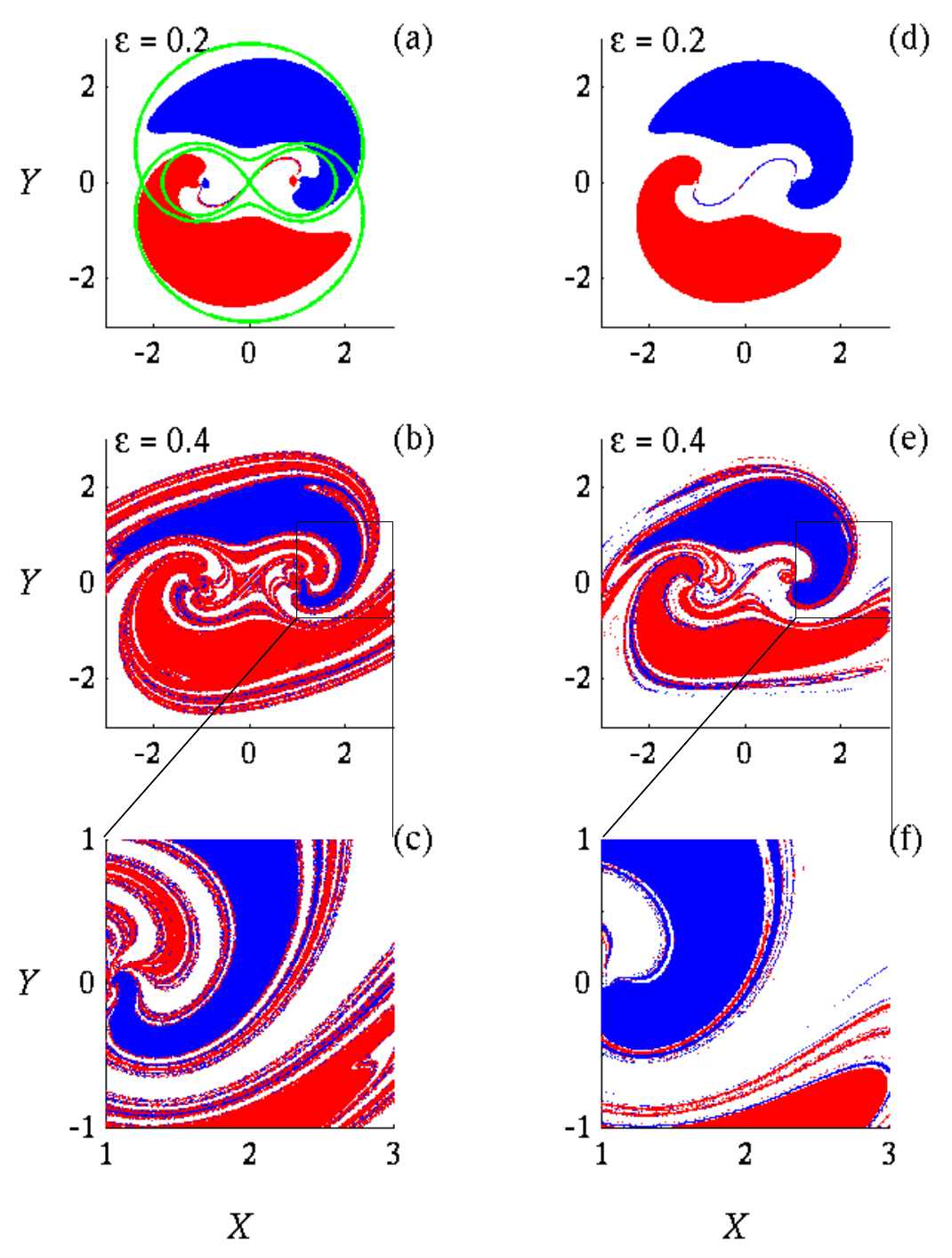}
        \caption{\baselineskip 13pt
        Counterpart of Fig.\ \ref{BasinsSt0.02} for $\St = 0.07$.  The differences between the perturbative   velocity field simulations in panels (a-c) and the exact potential flow simulations in panels (d-f) are now more noticeable  because $\vepsil$ is larger. The transition from smooth to fractal is in both cases in good agreement with the theoretical prediction  from Eq.\ (\ref{critere}).
        }
        \label{BasinsSt0.07}
\end{figure}

We have verified this by 
computing the basins of attraction numerically. Specifically, we plotted the initial $(X,Y)$ conditions of the trapped particles color-coded according to which of the two attractors they approach asymptotically.  Typical attraction basins computed through this procedure are shown in Fig.\ \ref{BasinsSt0.02} for $\St = 0.02$ and in Fig.\ \ref{BasinsSt0.07} for $\St = 0.07$;
these figures were generated using $2\times 10^5$ particles  initially uniformly distributed in a square region covering the vortices and released with initial velocity equal to the local fluid velocity. 
 For $\St = 0.02$,  
 Eq.\  (\ref{critere}) predicts the formation of a heteroclinic tangle  at $\varepsilon  \approx 0.15$. Indeed, Fig.\ \ref{BasinsSt0.02} shows that  the basin boundaries appear smooth for $\varepsilon =0.1$ but have filamentary characteristics for $\varepsilon=0.2$. In order to check the predictions of the perturbative model 
in Eq.\ (\ref{innerVf}),
we have also computed  the same attraction basins for particles advected by the exact 
four-vortex
potential flow. 
Even though  the detailed structure of the filaments is different, the overall shape of the basins obtained from the perturbative flow is close to the one obtained from the exact velocity field
even for relatively large $\varepsilon$
 (Figs.\ \ref{BasinsSt0.02}(b) and \ref{BasinsSt0.02}(e), respectively). Similarly, for  
$\St=0.07$ the theory predicts the formation of a heteroclinic tangle at  $\varepsilon  \approx 0.28$, in agreement with the basins shown in Fig.\ \ref{BasinsSt0.07}. As expected, for larger $\vepsil$'s---as used in Figs.\ \ref{BasinsSt0.07}(b) and \ref{BasinsSt0.07}(e)---more significant discrepancies appear between the perturbative and exact flow simulations.  In particular, the external heteroclinic orbit $\Sigma_3$ (considered in the next section) is closer to the vortices and may affect the particle dynamics, and this effect is not captured by the internal perturbative model.

In order to further validate the predictive power of Eq.\   (\ref{critere}) we have computed the fractal dimension of the basin boundaries for various $\varepsilon$'s and various Stokes numbers.  
The dimension can be computed efficiently and accurately from a sample of representative points in the boundary \citep{Grassberger1986}.
To generate a set of such points we applied bisection on a segment of line cutting the basin boundary.
Specifically,  to search for a point in the boundary
we 
randomly pick a pair of points in the  line segment 
 $x = 2.3$ and $-1\le y\le 1$,
 which is close to $\Sigma_2$, the first separatrix to break as the  
perturbation parameter  $\varepsilon$ 
increases.  The pair is discarded if both points 
are found to be in
the same basin of attraction. Otherwise  
we determine
the basin to which the midpoint of the segment joining that pair belongs. This allows us 
to form a pair of 
points closer to each other belonging to different basins.
The procedure is repeated until 
we obtain
 points belonging to different basins and at a distance from each other that does not exceed some pre-defined 
threshold
$2 d_{thr}$. 
This implies that the midpoint of the final pair is less than
$d_{{thr}}$-appart     
from  
the basin boundary
and hence serves as a good approximation to a point in the boundary.
After identifying
a few hundred 
such points   
using this algorithm, we 
applied
a method introduced 
in \cite{Grassberger1986} to compute 
the dimension. 
The method is based on the scaling $\langle  1/N_i (R) \rangle \sim R^{-D^{(1)}}$, where $N_i (R)$ is the number of sampled points within a ball of radius $R$ centered at the $i$-th point and $\langle \cdot  \rangle$ denotes the average over all $i$.
The dimension $D^{(1)}$ refers to the intersection set between the basin boundary and the initial line segment, while the dimension of the basin boundary in 2-dimensional portraits such as those in Figs.\ \ref{BasinsSt0.02} and  \ref{BasinsSt0.07} is simply $D^{(2)} = 1+D^{(1)}$.
The dimension of the basin boundary  in the full 4-dimensional phase space of the inertial particle dynamics
is   $D^{(4)} = 3+D^{(1)}$.

\begin{figure}
\vskip5cm
        \centering
     \includegraphics[width=0.65\textwidth]{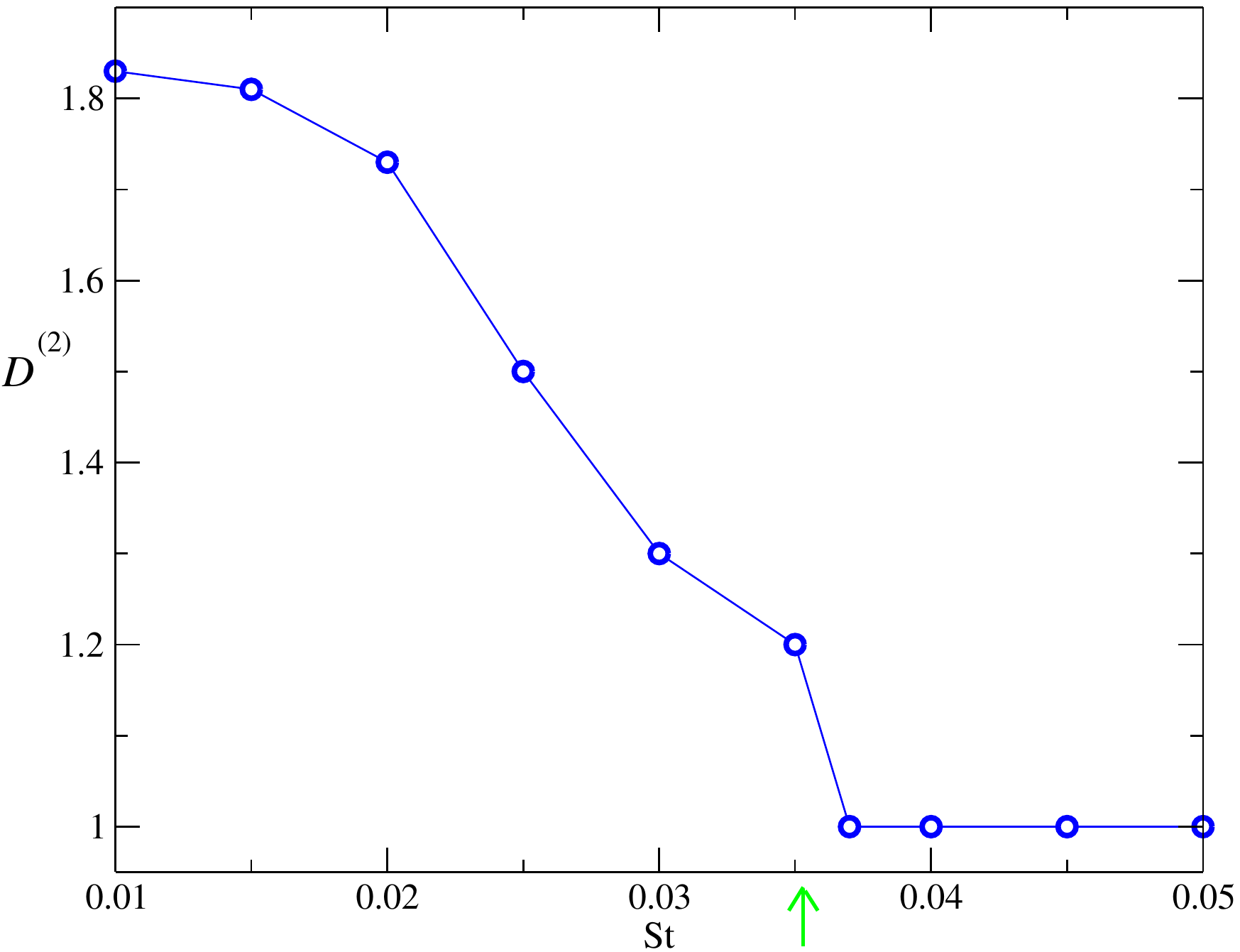}  
 \caption{\baselineskip 13pt
 Transition from smooth to fractal basin boundary as particle inertia is increased.
Basin boundary dimension as a function of the Stokes number for $\varepsilon=0.2$. 
The basin boundary is fractal when $\St \lessapprox \St_{c_2}\!= 0.0352$, in agreement with our theory (green arrow).  
The symbols correspond to numerical simulations and the continuous line is a reference to guide the eye.
} 
        \label{DvsStEpsi0.2}
\end{figure}

\begin{figure} 
      \centering
                \includegraphics[width=0.65\textwidth]{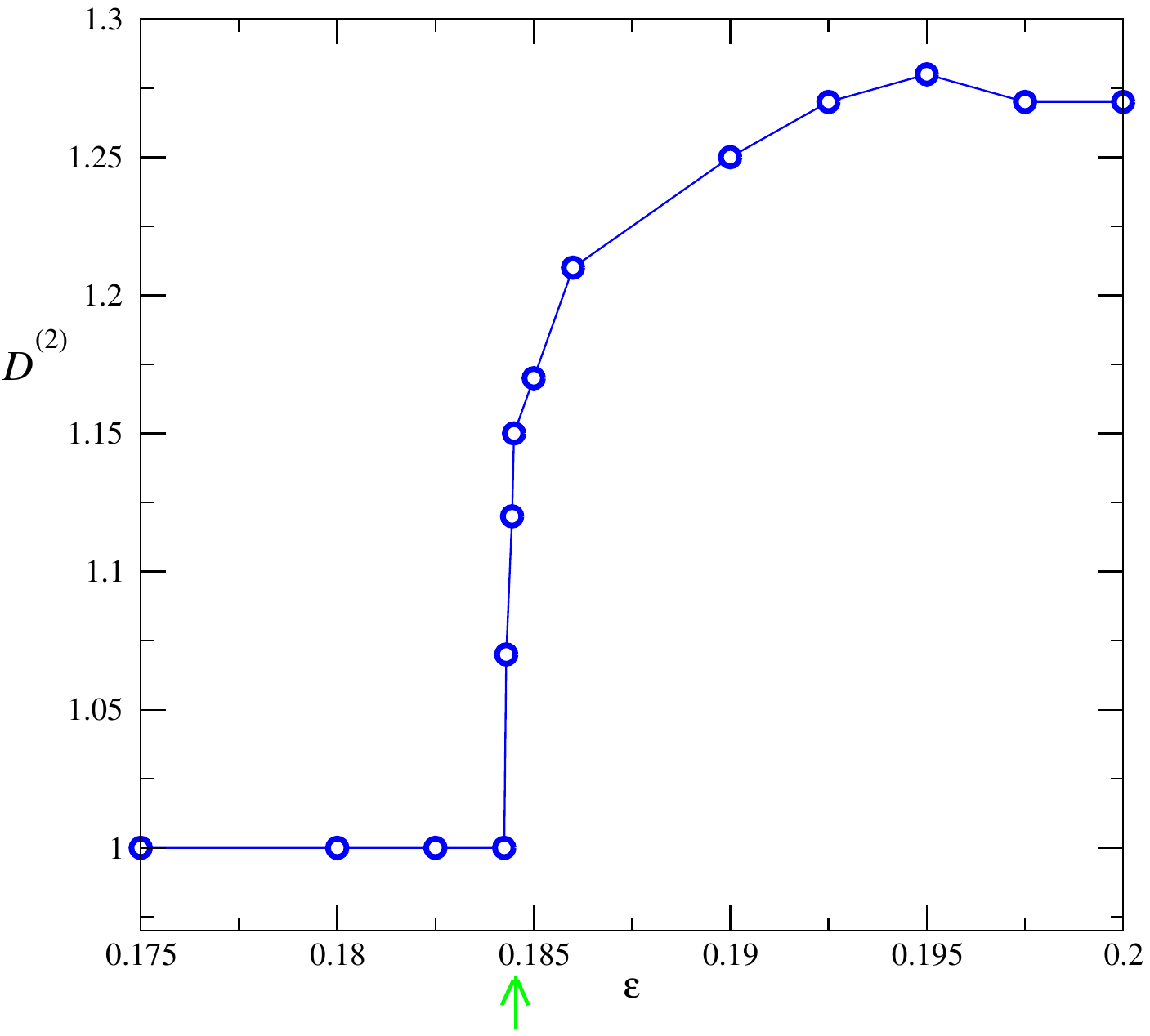} 
\caption{\baselineskip 13pt
Transition to fractal basin boundary as the perturbation of the flow by the wall is increased. The symbols correspond to numerical simulations of the basin boundary dimension as a function of $\varepsilon$ for St  =0.03. At $\varepsilon \approx  0.1846$  
the basin boundary ceases being smooth, in quantitative agreement with our theory (green arrow). 
The continuous line is a reference to guide the eye.} 
     \label{DvsEpsiAsymSt0.03}
\end{figure}

Figure \ref{DvsStEpsi0.2}  shows $D^{(2)}$
versus St for $\varepsilon=0.2$.  In this case, Eq.\ (\ref{critere}) predicts that   a heteroclinic tangle 
 exists when $\St  < 0.88 \, \varepsilon^2 \approx 0.035$. 
We indeed observe that the basin boundary is fractal ($D^{(2)} > 1$) when $\St \lessapprox 0.035$ and smooth ($D^{(2)} = 1$)  otherwise. Figure \ref{DvsEpsiAsymSt0.03} shows $D^{(2)}$ versus $\varepsilon$ for St = 0.03. In this case, Eq.\ (\ref{critere}) predicts that   a heteroclinic tangle exists when  $\varepsilon > (\mbox{St} / 0.88)^{1/2}  \approx 0.185$. 
This corresponds to the critical distance to the wall below which the particle dynamics becomes chaotic in the vicinity of the  separatrix $\Sigma_2$.
The numerical calculation shown in Fig.\ \ref{DvsEpsiAsymSt0.03}  confirms that 
the basin boundary  is
indeed smooth for $\varepsilon \lessapprox 0.185$ and fractal for larger $\varepsilon$. This is consistent with the expectation that the chaotic set around this (internal) separatrix gives rise to the fractal structure of the basin boundary.

Next, we consider the flow further away from the attractors and the (external) separatrix that exists between bounded and unbounded streamlines.

\section{Trapping of heavy particles from the open flow}   
 \label{sepS1S2}

The flow investigated in the previous sections is bounded by the heteroclinic orbit $\Sigma_3$ (Fig.\ \ref{ComparePsiAsym4VortEpsi0.25}(a), bold curve). This external separatrix is the boundary between the closed streamlines near the vortices and the open streamlines going to infinity.  When the Stokes number is sufficiently small, the velocity of  the particles is close to the local fluid velocity and hence the separatrix  $\Sigma_3$ also appears in the leading-order phase portrait of inertial particles. Yet, for any nonzero inertia, the  corresponding invariant manifolds associated with the saddle points $S_1$ and $S_2$ no longer coincide. Nevertheless,  as we show below, no particles from outside can cross the separatrix if the invariant manifolds split apart.  This is so because the invariant manifolds shield the flow region internal to $\Sigma_3$ through a mechanism analogous to the one described in Fig.\ \ref{SketchManifolds} for the separatrix $\Sigma_2$. Under these circumstances, particles released outside  $\Sigma_3$  will never reach the neighborhood of the vortices and will never be captured by the attractors investigated in Sec.\ \ref{secAsym}.  The scenarios in which the trapping of particles from the open flow occurs are investigated in this section. We show that, as in the case of the separatrix $\Sigma_2$, the  emergence of transverse intersections between the invariant manifolds is a necessary condition for particles to cross the separatrix $\Sigma_3$.

\subsection{Perturbative external fluid velocity field}

The  typical length and velocity scales of the flow near the separatrix $\Sigma_3$ are $L_0$ and $\Gamma/L_0$, respectively.  
Hence, we non-dimensionalize the streamfunction by using $L_0$ for lengths and $\Gamma/4\pi L_0$ for velocities. 
This non-dimensionalization is different from the one introduced in Sec.\ \ref{secAsym} for the internal flow.  In the analyses below we continue to use the same notation for the dynamical variables with respect to the new non-dimensionalization. To facilitate comparisons, however, in all figures we continue to use spatial coordinates normalized by $d_0$, as done in our analysis of the internal flow. 

From the external separatrix, to first approximation, each pair of vortices can be seen as a single vortex. Therefore, we make use of the reference frame $x'O'y'$ translating with respect to the laboratory frame $xOy$ at velocity $v_{{}_0}$,  which is the leading order of the velocity of  the vortex pair.
In this translating frame, the non-dimensional streamfunction of the flow induced by the vortex pair plus its mirror is $\psi_E(x',y') = ({\psi - v_{{}_0}  \, y'})/{\Gamma/4\pi}$, where $\psi(x,y,t)$ is the streamfunction of the flow observed in the laboratory frame. Still assuming that $\vepsi = d_0 / L_0 \ll 1$,  the streamfunction can be expanded as  \citep{Angilella2011}
\beq
\psi_E(x',y') = \psi_0(x',y') + \vepsi^2 \psi_c(x',y') \cos \frac{2 t} {\vepsi^2}+ \vepsi^2 \psi_s(x',y') \sin \frac{2 t}{\vepsi^2}   + O(\vepsi^4),
\label{psiext}
\eeq
where $ \psi_0(x',y')$ is the streamfunction of a simple dipole centered at (0,0) (i.e., a single vortex plus its mirror vortex) and the $\varepsilon^2$ terms express 
the fact
that in reality
we have vortex {\it pairs} and the resulting flow is unsteady.

\subsection{Crossing of the external separatrix}

The characteristic time of the flow close to the external separatrix is $\varepsilon^{-2}$ times larger than the characteristic time of the flow close to the internal separatrices. The Stokes number for heavy particle dynamics near $\Sigma_3$ is therefore equal to $\varepsilon^2 \St$,  
where $\St$ is the previously introduced particle Stokes number in the internal flow.  The equation of motion of the particles in the velocity field corresponding to the streamfunction (\ref{psiext}) then reads (removing the star superscripts for clarity and neglecting terms of order higher than two in the fluid velocity), 
\beq
\frac{d^2 \vec{X}_p}{dt^2} = \frac{1}{\vepsi^2 \St}\left(\vec V_{0}(\vec X_p)  + \varepsilon^2 \vec V_{2c}(\vec X_p) \cos \frac{2 t}{\vepsi^2} + \varepsilon^2 \vec V_{2s}(\vec X_p) \sin \frac{2 t}{\vepsi^2}  - \frac{d \vec{X}_p}{dt} \right),
\label{PtclMotionExtSep}
\eeq
where the velocity fields $\vec V_{0}$, $\vec V_{2c}$ and $\vec V_{2s}$  correspond to the streamfunctions $\psi_0$, $\psi_c$ and $\psi_s$ respectively.  Keeping $\St$ fixed and expanding the particle velocity in powers of $\vepsi$, we obtain
\begin{eqnarray}
\frac{d \vec{X}_p}{dt} =&& \!\!\! \vec V_{0}  - \St \, \varepsilon^2 \,  \vec V_{0} \cdot \nabla \vec V_{0}   
+ \varepsilon^2 \, (\vec V_{2c}   - 2 \St \vec V_{2s}  ) \cos \frac{2 t } {\vepsi^2}    \nonumber\\
&+& \varepsilon^2 \, (2\St \vec V_{2c}   + \vec V_{2s}  ) \sin \frac{2 t } {\vepsi^2}  + O(\vepsi^4, \vepsi^2 \St^2).
\label{rapidlyXpdot}
\end{eqnarray}
We therefore have a rapidly perturbed Hamiltonian system, with a perturbation frequency $\sim \varepsilon^{-2}$ \citep{Gelfreich1997}. 

One can always 
calculate the Melnikov function 
$M(t_0)$
representing 
the signed distance between the unstable and stable manifolds 
of
the saddle points $S_1$ and $S_2$, respectively. 
But in this rapidly perturbed  system  
the Melnikov function itself depends on $\vepsi$, which contrasts with the Melnikov functions of the internal separatrices.
Indeed, we obtain that
\begin{equation}
M(t_0) = - m \, \St  + A(\vepsi) \left( \sin \frac{2 t_0}{\vepsi^2} - 2 \St \cos \frac{2 t_0}{\vepsi^2}\right),
 \label{MelnikRapidly}
\end{equation}
where the constant $m$ 
  represents 
centrifuge effects and the amplitude $A(\vepsi)$ is the contribution of the unsteady perturbation due to the rotation of  
the
vortices around each other.
To the best of our knowledge, it has not been rigorously demonstrated that 
 simple
zeros in a 
Melnikov function of this form will 
 necessarily imply that the dynamics is chaotic \citep{Gelfreich1997}.
Nevertheless, the existence of such simple zeros of $M(t_0)$
guarantees that  particles  can cross  $\Sigma_3$ in both directions, 
and hence that a fraction of particles from the open flow can enter the closed component of the flow.
In contrast,  a negative sign for all $t_0$ 
in the
 Melnikov function 
indicates 
that particles released outside cannot enter.

The constant $m$ can be written as
\begin{equation}
m = \int_{-\infty}^\infty \left[\vec V_{0} \times ( \vec V_{0} \cdot \nabla \vec V_{0}) \right](\vec q(t)) dt,
\end{equation}
where
$\vec q(t)$ is a solution of the unperturbed system on the separatrix $\Sigma_3$, which we 
calculated
numerically, leading to $m \simeq 30.4$. 
The amplitude of the oscillating part  
reads
\begin{equation}\label{amplitude}
A(\vepsi) = \int_{-\infty}^\infty \left[ \vec V_{0} \times  \vec V_{2s} \right] (\vec q(t))\,\cos \frac{2 t}{\vepsi^2} dt - \int_{-\infty}^\infty \left[\vec V_{0} \times  \vec V_{2c}\right] (\vec q(t))\,\sin \frac{2 t}{\vepsi^2} dt,
\end{equation}
and was
computed numerically using a grid for $\vepsi \in [0,0.5]$. The result 
was
 then fitted with a combination of exponential and rational functions of $\vepsi$ 
as
\begin{equation}
A(\vepsi) \simeq \frac{e^{-\beta_3/\vepsi^2}}{\vepsi^2}(\beta_0 + \beta_2 \vepsi^2),
\end{equation}
where
$\beta_0 \simeq 23.6$, $\beta_2 \simeq 46.0$, and $\beta_3 \simeq 0.63$. Finally, by
imposing
that the 
oscillatory part
be 
smaller than the constant part of $M(t_0)$, we obtain a sufficient condition for  $\Sigma_3$ to be closed for particles released outside:
\beq
\St > \St_{c_3}(\vepsi) \equiv    \frac{e^{ - 0.63 / \vepsi^2}}{\vepsi^2} (0.78 - 1.51  \vepsi^2), 
\label{critS1S2}
\eeq
where the constants were replaced by their numerical values, and we made use of the fact that $|\sin(x)- 2 \St \cos(x)| \le (1+4 \St^2)^{1/2}$ for all $x$. 
In addition, to obtain a simpler criterion, we have assumed that $\St^2$ is small compared to $1$.
Because integrals have been fitted, Eq.\ (\ref{critS1S2}) is a partially numerical criterion rather than a purely analytical one. Nevertheless, this formula is very useful to predict trapping from the open flow, and is used in the next section to construct the complete trapping diagram of particle dynamics in the vortical flow.

\subsection{Trapping diagram}

\begin{figure}
        \centering
                \includegraphics[width=0.85\textwidth]{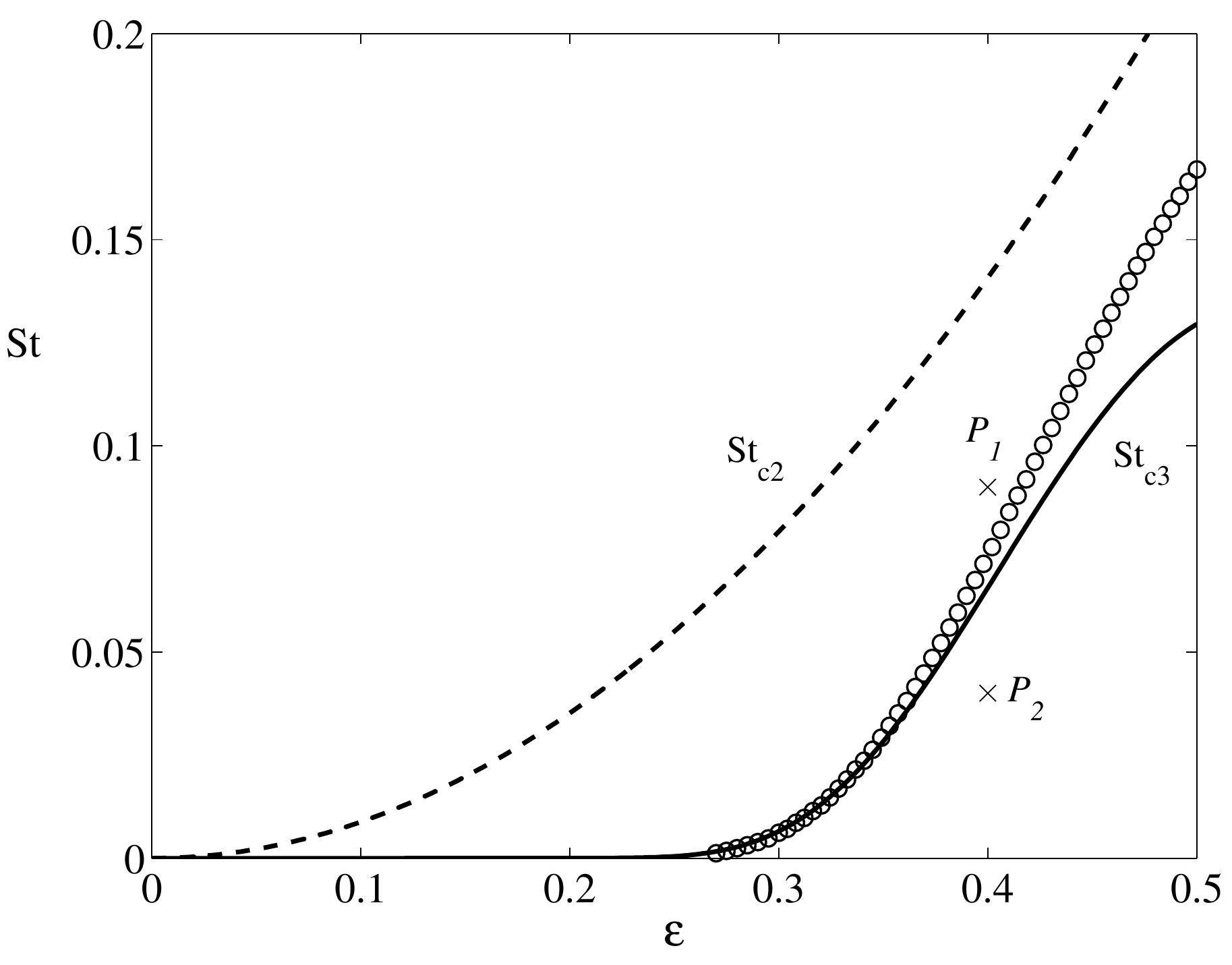} 
 \caption{\baselineskip 13pt
Global trapping diagram for heavy particles in the absence of gravity.
The dashed curve is the
theoretical critical Stokes number $\St_{c_2}$  for the opening of the internal separatrix $\Sigma_2$ (Eq.\ (\ref{critere})), whereas the continuous curve is the theoretical critical Stokes number $\St_{c_3}$ for the opening of the external separatrix $\Sigma_3$  (Eq.\ (\ref{critS1S2})). These curves define three regions according to whether external particles can cross the corresponding separatrix. 
Circles represent a numerical verification of $\St_{c_3}$ based on simulations of the exact potential flow. Parameter points $(\vepsil,\St)$ above the circles (such as $P_1$) correspond to scenarios in which particles released in the open component of  the flow do not cross $\Sigma_3$, while  points below the circles (such as $P_2$) correspond to scenarios in which a fraction of them do cross  inside. 
}

  \label{StEpsiXFIG}
\end{figure}

Figure \ref{StEpsiXFIG} shows in the $(\vepsi,\St)$ plane both the critical Stokes number $\St_{c_3}$ for the  breakdown of the external separatrix $\Sigma_3$  (defined by the converse of Eq.\ (\ref{critS1S2})) and the  critical Stokes number $\St_{c_2}$  for the breakdown of the internal separatrix $\Sigma_2$ (defined by the converse of Eq.\ (\ref{critere})).  Above curve $\St_{c_2}(\vepsi)$  the separatrix  $\Sigma_2$ is ``closed" (i.e., the stable and unstable manifolds do not intersect
transversally,
and  cannot be crossed from the outside) and below this curve the separatrix  $\Sigma_2$ is ``open"  (i.e., the manifolds intersect transversely and a chaotic set is formed near the separatrix). A similar characterization applies to the curve $\St_{c_3}(\vepsi)$ with respect to the separatrix $\Sigma_3$.  Because $\St_{c_2} >\St_{c_3}$ for  $\vepsi>0$, the opening of $\Sigma_3$ implies that $\Sigma_2$ is also open. 
This does not mean  
that particles released outside and crossing $\Sigma_3$ will 
necessarily
 cross $\Sigma_2$, since the tangles around each separatrix do not necessarily overlap. However,
  if trapping occurs for particles released outside $\Sigma_3$,  then the parameters must be in the region defined by $\St < \St_{c_3}(\vepsi)$.  That is, being in this region, and hence having both separatices open,  is a {\it necessary} condition for particle trapping from the open flow.

In order to check  the
theoretically-predicted
 trapping diagram, we have implemented numerical simulations using the exact four-vortex potential velocity field $\vec V_f$ for the fluid
and the  
dynamical equation 
of 
the particles in the  laboratory frame (that is, Eq.\ (\ref{PtclMotion}) 
for $\vec W_f$ replaced by  $\vec V_f$  and
 without  the centrifugal and Coriolis forces).
 For given values of $\vepsi$ and $\St$,
 we considered
 particles 
 released far ahead of the vortices, outside $\Sigma_3$ 
 and
 near the wall. They are
 driven by the flow toward the vicinity of $S_1$ and then around the vortex pair near $\Sigma_3$, 
independently of
 the detailed shape of the initial distribution
 of particles.
Trajectories were  computed for a large number of turnover times, and the number  $N(\St)$ of particles 
 crossing
inside  $\Sigma_3$ during 
this period of time
was then counted.
The 
critical
Stokes number was
estimated numerically in these simulations
using
a bisection procedure 
applied to $N(\St)$,  
with the  process  
 terminated
when the difference in $\St$ for crossing or not crossing the separatrix fell
below a certain threshold. 
The result is plotted in Fig.\  \ref{StEpsiXFIG} (circles): no particle released 
in the open flow
is observed to cross $\Sigma_3$ when  $\St$ is below the circles. We 
note
that this numerical curve agrees with the theoretical value of  $\St_{c_3}$
 in Eq.\ (\ref{critS1S2}) up to 
 $\vepsi  \simeq 0.4$. For larger $\vepsi$, that is when vortices are 
 closer
 to the wall, the perturbative theory underestimates the critical Stokes number. This might be due to the fact that the wall-induced perturbation is underestimated by the
perturbative velocity field there.

\begin{figure} 
      \centering
                \includegraphics[width=0.70\textwidth]{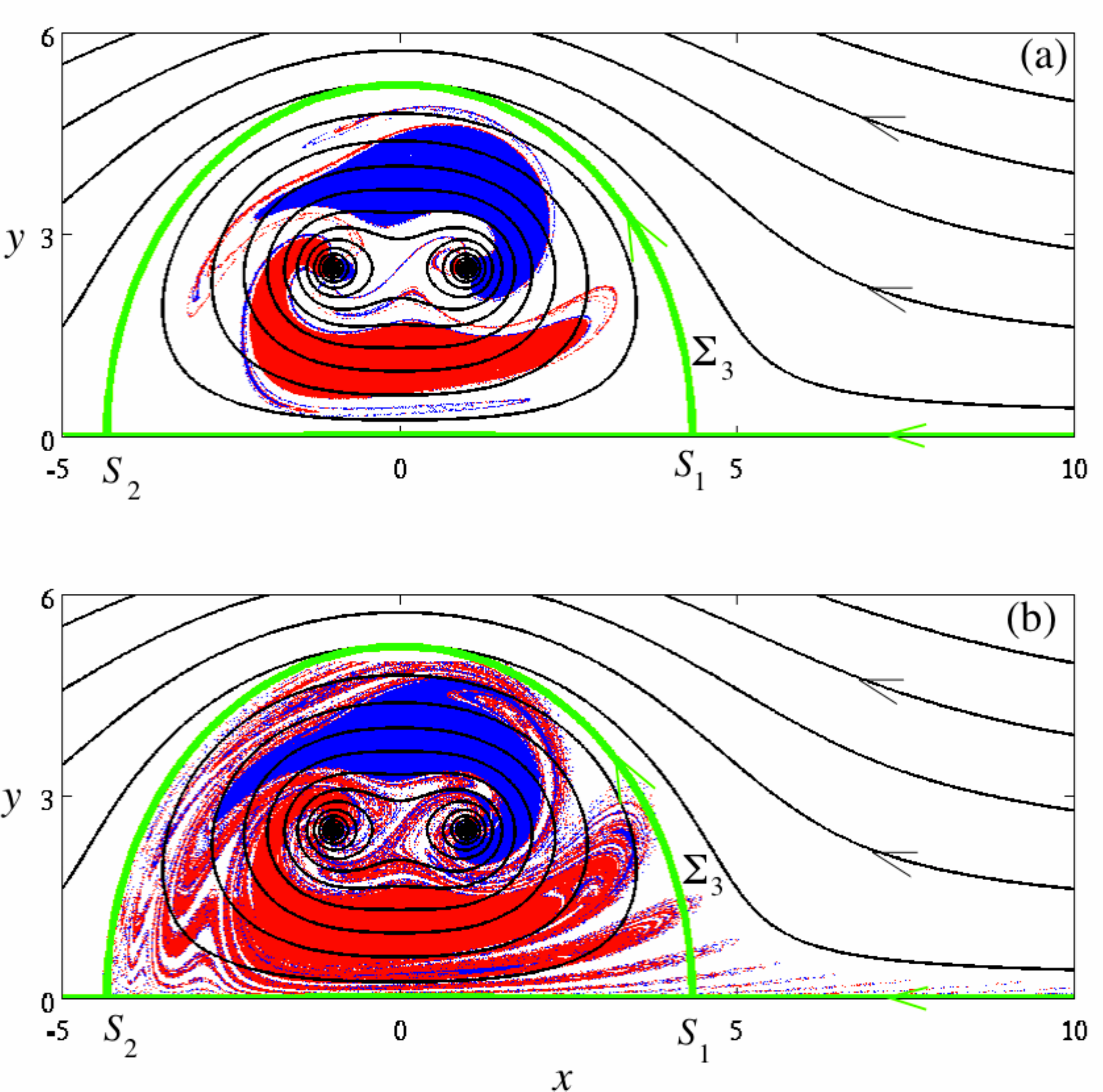} 
\caption{\baselineskip 13pt
Bounded versus unbounded basins of attraction. 
Red and blue represent the basins of attraction
 for
(a) point $P_1$  and (b) point $P_2$  in the diagram of Fig.\ \ref{StEpsiXFIG}, obtained using 
simulations of
the exact potential flow. The continuous lines represent instantaneous streamlines. In contrast with panel (a), the 
basins
of attraction in panel (b) extend 
outside the external separatrix $\Sigma_3$, demonstrating the existence of trapping from the open flow.}
\label{BassinsAetB}   
\end{figure}

To further  
examine the validity of
the trapping diagram, two simulations---correspond-ing to the parameters $P_1$ ($\vepsi = 0.4, \St = 0.09$) and $P_2$ ($\vepsi = 0.4, \St = 0.04$)  
in the  diagram of Fig.\ \ref{StEpsiXFIG}---have been carried out   
for the exact potential velocity field induced by the four vortices. 
The initial positions of the vortices are $(\pm 1, \pm 1/\vepsi)$, and particles are released 
in the 
rectangle $[-5, 10]\times [0, 5]$, 
extending to the open portion of the flow ahead of the vortices  and meant to 
detect whether the attraction basins reach outside $\Sigma_3$. 
 The basins of attraction defined by these initial conditions are shown in  Fig.\ \ref{BassinsAetB}.
 The basins of attraction 
  extend outside the separatrix $\Sigma_3$ in the case of $P_2$, and  
  are 
contained within it in the case of $P_1$.  
This is in accordance with our theoretical predictions that
 trapping from the open flow would be possible for $P_2$  
but not  for $P_1$, 
 which is also confirmed by direct simulations of  both the perturbative and the exact potential velocity field. In the case of $P_2$, this means that a 
fraction of the particles crossing $\Sigma_3$ can
also cross $\Sigma_2$ and approach the attractors. 
We deduce that this mechanism underlies the trapping of heavy particles in 
the leapfrogging open vortical flow
observed in the previous numerical study of  \cite{Vilela2007}. 
Figure \ref{BassinsAetB}
also suggests that the probability of getting trapped from the outside is small, since the measure of the external portion of the basins is small compared to the volume of the tested region. However, the figure also indicates that this probability is much larger for particles released near the wall.

\section{Robustness of trapping}
\label{simulns}

In the previous sections we have shown that our perturbative analysis successfully describes trapping of heavy particles in the exact potential flow of a vortex pair and its specular image. It is natural to   
consider
whether trapping from the open flow is a robust phenomenon
in the presence of other factors that 
might
not be negligible  
in realistic situations.
Specifically,
we show below that the
trapping 
 of heavy particles from the open flow
 also occurs when the  
particles are subject to 
gravity, in the presence of noise, and when the 
potential flow is replaced by a viscous flow obtained from 
direct
 simulations of the Navier-Stokes equations.
 For 
 clarity,
 we consider each of  these three 
effects separately.

\subsection{Effect of gravity}
\label{secgravity}

We use $\theta$  to denote the angle between the gravitational field $\vec g$ and the axis perpendicular to the wall, such that  $\vec g = g(\sin \theta \, \hat {\vec x} - \cos\theta \, \hat {\vec y}$), 
and assume that the particles have a small but nonzero 
 settling velocity $\vec g \, \tau_p$.  In the case of non-vertical walls, the settling velocity is set to zero in a thin layer above the wall in order to account for the finite size of the particles in the particle-wall interactions (e.g. lubrication forces) and prevent particles from crossing the wall in the simulations. In the previous sections, which included no gravity term, this precaution had not been applied since inertia alone cannot lead to the crossing of the symmetry line for small Stokes numbers (i.e. there is no inertial impaction). 
 
Applying
the same method 
used in  Sec.\ \ref{partmotion} \citep{IJzermans2006}, 
 it can be verified that
 attracting 
 points still exist in the presence of gravity 
 provided that the settling velocity is not too large.
 Moreover,  it can be shown that in this regime the
opening of the internal separatrix  $\Sigma_2$ is only weakly influenced by gravity and still occurs before the opening of the external separatrix, $\Sigma_3$.
  The question 
  then
  is whether particles 
  released
  in the open 
   flow can
  cross into the closed component and
  be captured by the attracting points or possibly by a new attractor.
  To address this question we investigate 
the opening of
$\Sigma_3$ in the presence of gravity.

The   
equation of motion
 for a heavy particle  
 in the presence of
  gravity is obtained by adding the non-dimensional  weight force to the drag term. 
  Because we focus on the crossing of the 
  external 
separatrix $\Sigma_3$, it is convenient to use the external units $V_0 = \Gamma/4 \pi L_0$ (for velocities) and $L_0$ (for lengths) already used in the preceding section. This leads to
\beq
  \vepsi^2 \St \frac{d^2{\mathbf X_p}}{dt^2}= {\mathbf V}_f - \frac{d{\mathbf X}_p}{dt} + \widetilde{V}_T \hat{{\mathbf g}},
\label{ddotXpgravity}
\eeq 
where 
 $\hat{{\mathbf g}}$ is the  unit vector in the direction of gravity,
${\mathbf V}_f$ is the non-dimensional fluid velocity corresponding to the
streamfunction (\ref{psiext}), and $\widetilde{V}_T = g \tau_p / V_0$ 
is the non-dimensional 
free-fall terminal 
particle
velocity 
in still fluid. 
 To express 
 that the settling velocity, although small, is
sufficiently large to compete against the inertia term  (i.e., that the
gravity and inertia terms have the same (small) order of magnitude), 
we set $\widetilde {V}_T = \vepsil^2 \bar V_T$, where $\bar V_T$ is assumed to be of order unity. Then,  
expanding the particle velocity in powers of $\vepsi$ leads to Eq.\ (\ref{rapidlyXpdot}) with an extra additive
term $\vepsil^2 \bar V_T \hat{{\mathbf g}} $. This is again a rapidly perturbed 
Hamiltonian 
system with the 
same 
leading order
 as  Eq.\ (\ref{rapidlyXpdot}), but with a different perturbation.  
The gravity term
 results in an additive constant term in the
 Melnikov function:  
\beq
M_g(t_0) = M(t_0) + \bar V_T \Big( \sin \theta \, \int_{-\infty}^{\infty} \frac{\dr \psi_0}{\dr x'} dt
- \cos \theta \,  \int_{-\infty}^{\infty} \frac{\dr \psi_0}{\dr y'}  \, dt   \Big),
\eeq
where $M_g(t_0)$ denotes the Melnikov function in the presence of gravity and
$M(t_0)$ is the gravity-free Melnikov function
 given by Eq.\ (\ref{MelnikRapidly}). The first   integral
 in this equation
is
the difference   
$q_y(-\infty)-q_y(+\infty)$  for a point $\mathbf q(t)$
moving
 on $\Sigma_3$, and is equal to zero. The second 
 integral
is equal to $q_x(+\infty)-q_x(-\infty) = -2 \sqrt 3$. We finally obtain
\beq
M_g(t_0) =  {2 \sqrt 3 \bar V_T \cos \theta - m \, \St}   +  A(\vepsi)
\left(\sin \frac{2 t_0} {\vepsi^2} - 2 \St \cos \frac{2 t_0} {\vepsi^2} \right),
\label{Mg}
\eeq
where $A(\vepsi)$ is  defined in Eq.\ (\ref{amplitude}).

\vskip2mm
{\em Non-vertical wall}.
We first
assume that $-\pi/2 < \theta < \pi/2$, so that  gravity 
pulls the 
particles 
toward
the wall (the limit 
case  $\theta = 
-\pi/2$ 
is
discussed 
below).
It is immediate from 
Eq.\ (\ref{Mg}) that the constant term due to gravity,
$2 \sqrt 3 \bar V_T \cos \theta $, is positive  
and hence opposes the constant term due to centrifugal effects, $- m \, \St $, which is negative.
The last term, which is not constant, is a consequence of the unsteady perturbation due to the rotation of the vortex pair.
Three kinds of behavior therefore appear:

\vskip2mm
(i)   $M_g(t_0) < 0$ for all $t_0$: centrifugal effects dominate over both gravity and unsteadiness. The unstable manifold $W^u$  
of the
 perturbed hyperbolic-saddle point near $S_1$ 
  wraps around 
the stable  manifold
  $W^s$ 
  of
 the hyperbolic-saddle point near $S_2$. 
 The dynamics is regular, 
 and
 particles released outside cannot enter. Particles released inside sufficiently close to $\Sigma_3$ will spiral out. 
 This happens when
\beq
\St > \frac{2 \sqrt 3}{m}  \frac{V_T}{\vepsi^3}\cos \theta +\St_{c_3}(\vepsi) \equiv \St_{c_3}^+(\vepsi,V_T),
\label{RegulClosed}
\eeq
where $\St_{c_3}(\vepsi)$ is the gravity-free critical Stokes number given in Eq.\ (\ref{critS1S2}).
\vskip2mm

(ii)   $M_g(t_0) > 0$ for all $t_0$: gravity dominates over both centrifugal effects  and unsteadiness. The 
manifold $W^s$ now wraps around $W^u$.
The dynamics is regular,  
but
a fraction of the
particles released outside can spiral in.
This happens when
\beq
\St < \frac{2 \sqrt 3}{m}   \frac{V_T}{\vepsi^3}\cos \theta - \St_{c_3}(\vepsi)\equiv \St_{c_3}^-(\vepsi,V_T).
\label{RegulOpen}
\eeq

(iii)   $M_g(t_0)$ has simple zeros: 
due to the unsteadiness of the flow,
a chaotic saddle may exist  
in the vicinity of $\Sigma_3$ (see discussion following Eq.\ (\ref{MelnikRapidly})). 
In either case, the separatrix $\Sigma_3$ is necessarily permeable in both directions.
In particular, a fraction of the particles released outside penetrate inside after wandering in the
heteroclinic tangle.
This happens when
 \beq
 \St_{c_3}^-(\vepsi,V_T) < \St <  \St_{c_3}^+(\vepsi,V_T).
\label{ChaoticOpen}
 \eeq
In the formulae above, the settling velocity $\bar V_T$ has been replaced by $V_T/\vepsi^3$, where $V_T = g \tau_p / d_0 \Omega_0$ is the settling velocity in the unit system $(d_0,\Omega_0)$, which are the units  used 
in the
numerical simulations throughout this paper. 
Note
that $V_T$ must be  $O(\vepsi^3)$ for these asymptotic calculations  
 to be valid.

\begin{figure}
\centering
   \includegraphics[width=0.85\textwidth]{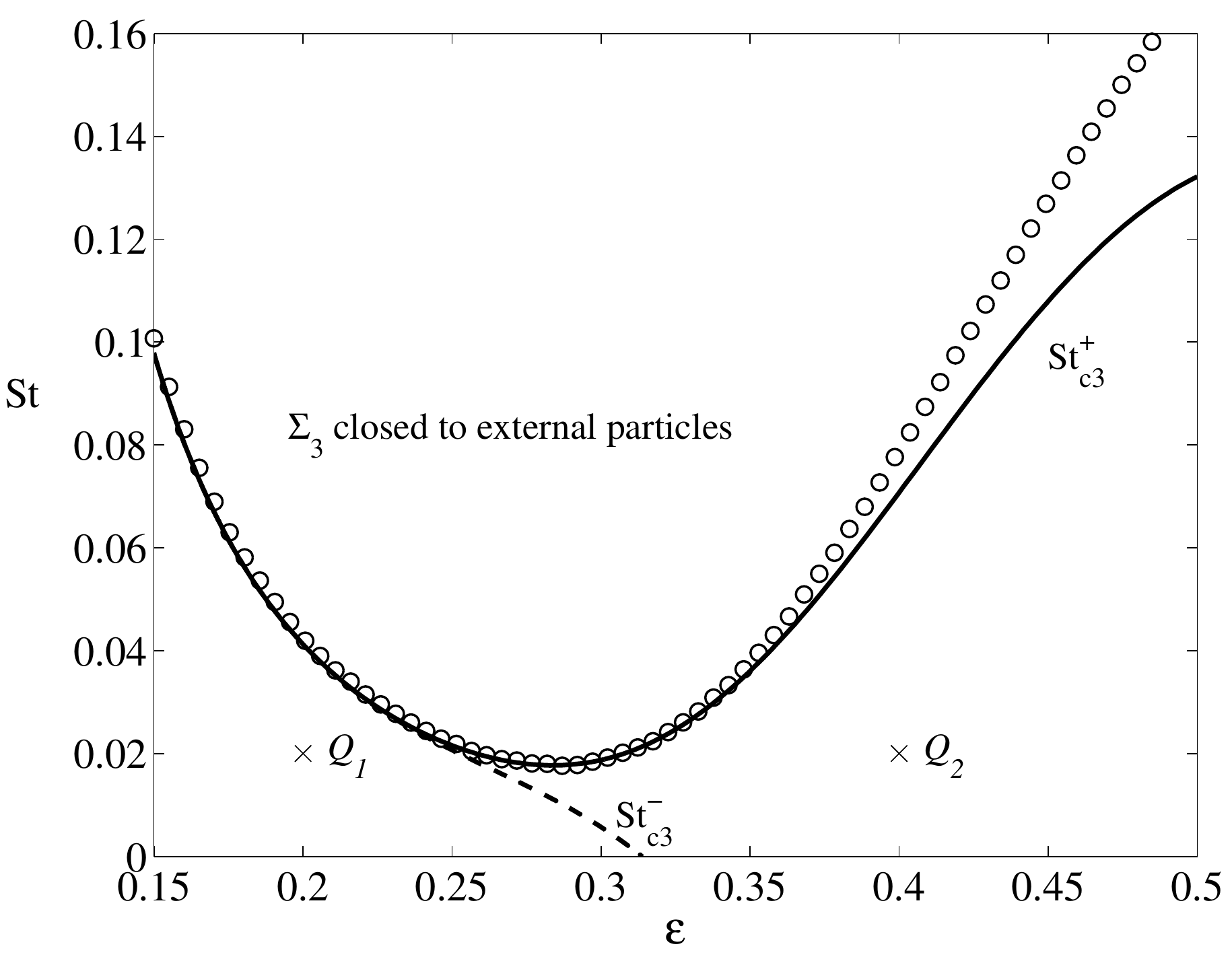}
\caption{\baselineskip 13pt
Trapping diagram for heavy particles in the presence of gravity for a horizontal wall.
The solid and dashed curves correspond to the theoretical critical Stokes numbers $\St_{c_3}^+$ (Eq.\
 (\ref{RegulClosed})) and $\St_{c_3}^-$ (Eq.\
 (\ref{RegulOpen})), respectively. These curves separate three possible behaviors determined by the Melnikov function  associated with $\Sigma_3$, in which particles can spiral out ($\St> \St_{c_3}^+$), spiral in (such as for point $Q_1$), or go both in and out (such as for point $Q_2$)  across this separatrix.
Circles correspond to a numerical verification of $\St_{c_3}^+$.  The diagram was generated using the choice $V_T = 0.003$ for the non-dimensional settling velocity. 
}
\label{diagramgravity} 
\end{figure}

Figure \ref{diagramgravity} summarizes our numerical verification of the theoretical predictions in Eqs.\
 (\ref{RegulClosed})-(\ref{ChaoticOpen}) for
$V_T = 0.003$ and $\theta = 0$ (horizontal wall). 
In this figure, which was generated employing the same method used to generate Fig.\ \ref{StEpsiXFIG}, we
show all three domains defined by Eqs.\ (\ref{RegulClosed})-(\ref{ChaoticOpen}) and use
circles to represent the parameters $(\vepsi,\St)$ above which no particle is observed to cross $\Sigma_3$. 
The predictions are in good agreement with the numerical results up to $\vepsi \simeq 0.4$, despite the fact that the numerical calculations were based on using the exact 
four-vortex
potential flow whereas the theoretical predictions were based on the external perturbative velocity field (\ref{psiext}).
These results  
reveal 
a new regime, where gravity can cause  particles released outside to 
 enter the region of closed streamlines 
without exhibiting (transiently) chaotic dynamics
 (case (ii)).

\begin{figure}
\centering
   \includegraphics[width=0.65\textwidth]{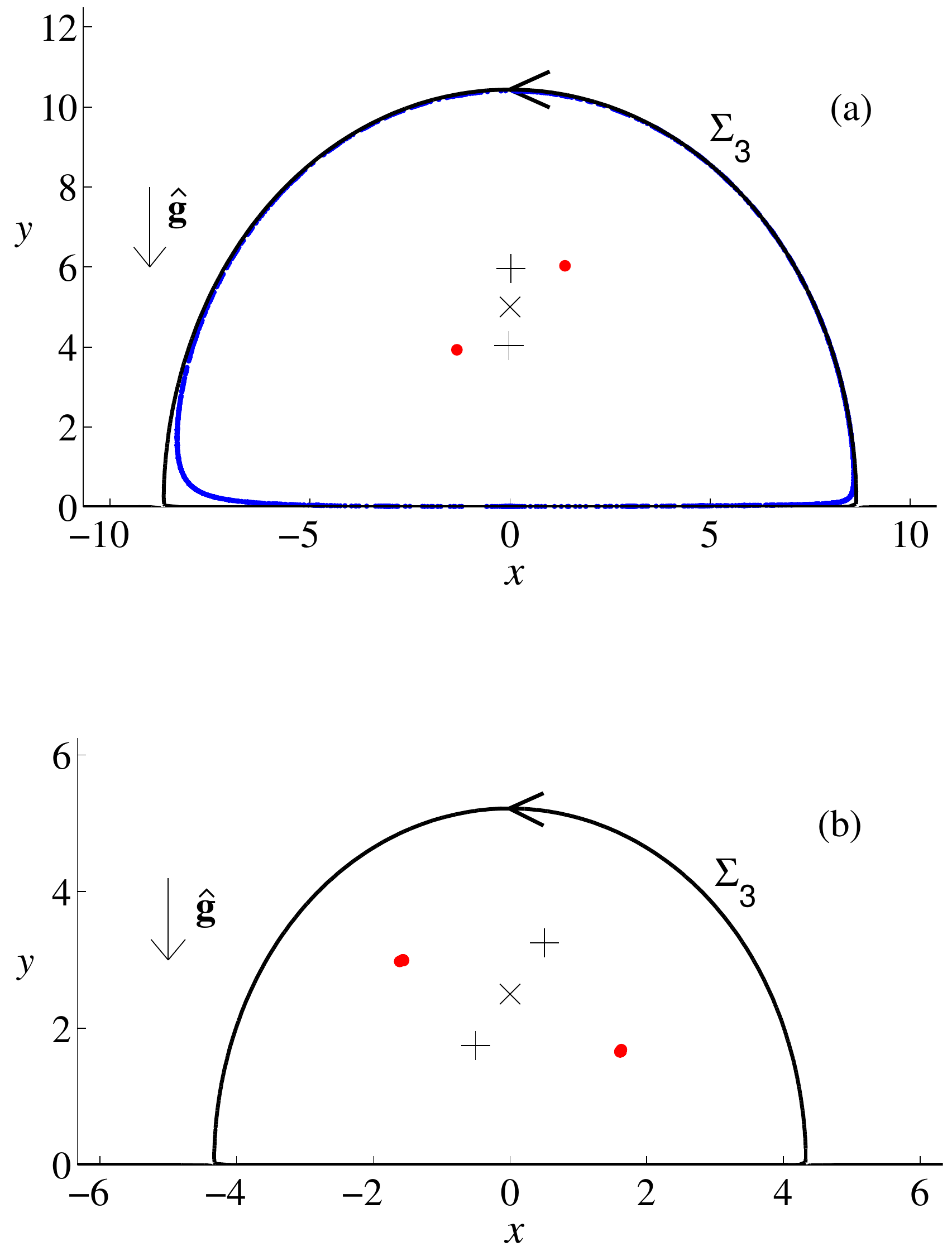}
\caption{\baselineskip 13pt
Attractors in the presence of gravity for a horizontal wall.
Colors indicate snapshots of the attracting sets for  (a) point $Q_1$  and (b) point $Q_2$  in the diagram of Fig.\ \ref{diagramgravity}, where
the plus symbols indicate the vortices at the same instant. The attracting sets were traced by evolving for a long period of time particles released in the open flow near the wall 
and  particles released in the closed flow covering the vortices.
In case $Q_1$, particles from the open flow cross the separatrix $\Sigma_3$ 
 under the sole effect of gravity and converge toward a limit cycle (blue) 
 right inside the separatrix; such particles cannot reach the point attractors (red) in the neighborhood of the vortices.
  Particles released inside the closed component of the flow can either converge to the limit cycle or be captured by the point attractors. 
In case $Q_2$,  a heteroclinic tangle exists near $\Sigma_3$ and there is no limit cycle.  The point attractors (red) can now trap not only particles from the closed flow but also a fraction of the particles from the open flow. The simulations were performed using  the exact potential flow.
}
\label{RunsGravity} 
\end{figure}

We observe that in this regime
particles 
can be permanently trapped
by a limit  cycle located inside 
but 
near $\Sigma_3$. 
 Figure
 \ref{RunsGravity} shows two simulations,  corresponding 
 respectively
 to the parameters $Q_1$ and $Q_2$ of the 
 trapping
 diagram of Fig.\ \ref{diagramgravity}.
 In each simulation we identified the respective attracting sets for
particles released both in the open
and in the closed flow by evolving their trajectories for a long period of time. 
  In the case of $Q_1$,  
    we observe that
   a fraction of the
 particles from the open flow penetrate into the 
    closed component
under the sole effect of gravity,  as predicted by our theory.
 These particles are trapped by  
 the
 limit cycle  
 near
 $\Sigma_3$ (and  hence  cannot reach the  
 attracting
  points
  near the vortices). This   limit  cycle exists 
  due to   the combined effect of gravity and inertia, and also because particles have finite size and are allowed to slip along the wall  (i.e.,   they do not stick to it). As described above,  the settling velocity is set to zero in a thin layer above the wall ($0 < y < \delta$).
This  allows resuspension, which is a key ingredient for the existence of this limit cycle. The exact value of the (small) thickness of the layer is of no importance for the existence of the limit cycle, but it affects its shape slightly (we used $\delta = 0.03$ in the computations of Fig.\ \ref{RunsGravity}). We have checked that, by decreasing the layer thickness $\delta$, the limit cycle passes closer to the right-side stagnation point, in agreement with the fact that the smaller the thickness, the later  the resuspension of the particles will be during their motion along the wall.  

Particles from the closed flow, however, can either
    be driven toward the limit cycle or spiral inside and be captured by the attracting points.
In the case of $Q_2$, 
on the other hand,
the limit cycle no longer exists and 
particles from both
 the closed and the open flow
are observed to approach the attracting points.
Moreover, 
plots of particle clouds at intermediate times 
(not shown)
confirm
that in this case
a heteroclinic tangle exists near $\Sigma_3$, as expected from the Melnikov analysis.

\begin{figure}
  \centering
 \includegraphics[width=0.65\textwidth]{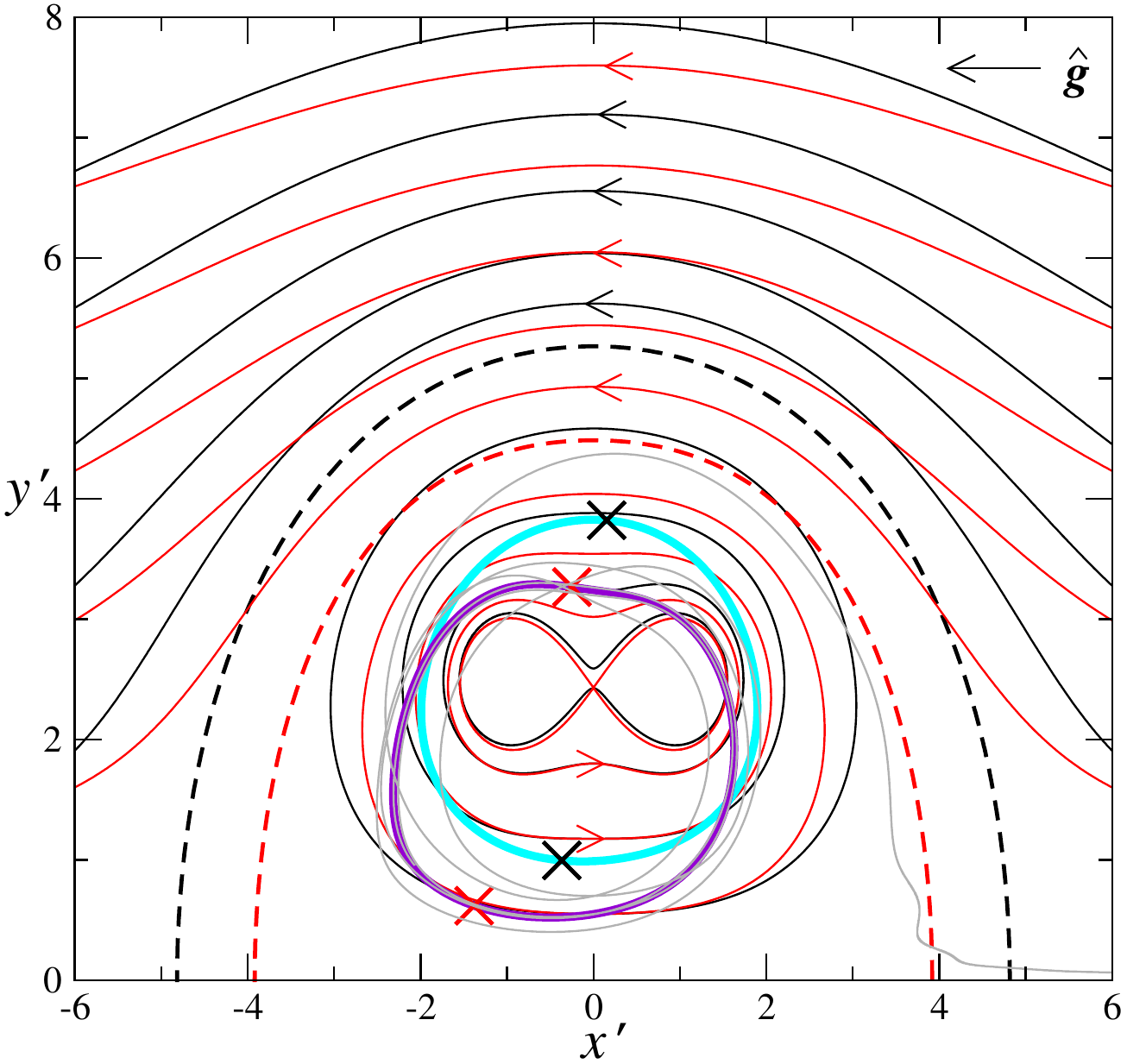}
\caption{\baselineskip 13pt                                
Particle trapping in the presence of gravity for a vertical wall. 
Snapshot of the 
streamlines of both
the fluid flow (continuous black lines) and 
the effective (``particle'') streamfunction 
 corresponding 
to the leading-order
velocity field of heavy particles under gravity (continuous red lines). 
Also shown are the
corresponding instantaneous separatrices (dashed lines) and the attracting points  
in the absence (black cross symbols) and presence (red cross symbols) of gravity at the same instant. The projections of the orbits of the attracting points into the physical space are represented in cyan and violet, respectively. 
The grey line shows the trajectory of a representative particle from the open flow  captured by a point attractor.
In these simulations we used  $\varepsilon=0.4$,  $\St=0.006$, and 
(for the presence of gravity) $V_{T}=0.28$.
The trapping of heavy particles from the open flow is observed under these conditions, and thus
persists even in the presence of a strong gravitational field
along the average direction of the fluid (in the reference frame of the center of vorticity). 
}
\label{streamgravity} 
\end{figure}

\vskip2mm
{\em Vertical wall}. In the limiting case $\theta = - \pi/2$,   vortices   
move
upward 
with respect to the laboratory frame,
and any heavy particle trapped in their neighborhood
would   be carried 
against the mean velocity of the fluid 
{\it and} against gravity 
instead of settling down. 
By considering
Eq.\ (\ref{Mg}) 
with this choice of angle,
we observe that
 gravity does not 
alter
 the Melnikov
function in this case. Small settling velocities, as considered so far, are therefore unable to affect significantly the dynamics 
of the particles
in the vicinity of the separatrix. 
In contrast, if the settling velocity is of the order of $V_0$, that is $\widetilde{V}_T = O(1)$ instead of  $O(\varepsilon^2)$ in
Eq.\ (\ref{ddotXpgravity}), then the particle velocity can be expanded as
\begin{equation}
\frac{d{\mathbf X}_p}{dt} = \vec V_f^0 - \widetilde{V}_T \hat{\vec x} + O(\vepsi^2).
\end{equation}
The leading-order particle dynamics now corresponds to the ``particle" streamfunction $\psi_p = \psi_0 - y \, \widetilde{V}_T$, which has been widely used as an elementary sedimentation model (see, for example, \cite{Stommel1949}).
 One can easily check that this streamfunction has the same general form as $\psi_0$, corresponding to a dipole with open streamlines  flowing around 
a closed region,
 but with a separatrix $\Sigma_3'$ 
smaller  than
 $\Sigma_3$.
 The separatrix $\Sigma_3'$
joins
two hyperbolic points, $S_1'$ and $S_2'$,
located at $x' = \pm [(6-\widetilde{V}_T)/(2+\widetilde{V}_T)]^{1/2}$ instead of $\pm \sqrt 3$ (in the external system of units). 
As long as $\widetilde{V}_T$ is not too large,
this 
structure will exist and the conditions leading to a heteroclinic tangle near $\Sigma_3'$ can be derived by a Melnikov analysis similar to the one described in the gravity-free 
case: one
 just needs to re-calculate the unperturbed trajectory $\vec q(t)$ on $\Sigma_3'$. To check  
 that
 attracting
 points still exist,  we have performed 
 simulations 
 in the case $\widetilde{V}_T = 0.28$, $\St = 0.006$ and
$\vepsi = 0.4$ (Fig.\ \ref{streamgravity}).    
The simulations confirm that
such
points do exist and that they capture particles coming
from the open 
portion of the flow.

\subsection{Effect of viscosity}

The calculations described in the above sections concern flows whose velocity fields are determined on the basis of the inviscid fluid approximation. That is, viscosity was assumed to be important only at scales comparable to or smaller than the particle diameter. However, when the flow Reynolds number is
only moderately large, viscosity is expected to play an important role also at scales of the order of the distance between vortices. At those scales it  leads to vortex merging, which eventually destroys the co-rotating vortex pairs.
Vortex merging starts when, due to viscous diffusion, the linear size of the vortex cores reaches a critical value of the order of the initial distance 
between the vortices (see, for example, \cite{Cerretelli2003} and references therein, or  \cite{Carton2002} for vortex merging in  an external strain flow).  We hypothesize that, if the time scale of viscous diffusion is much larger than the turnover time of the vortices, trapping of particles will occur as predicted by the potential flow theory (although only until vortex coalescence takes place).
 
To 
test
this hypothesis,
 we have performed a series of numerical simulations of the two-dimensional Navier-Stokes equations with an initial vortex pair parallel to the $x$-axis,  composed of two identical Lamb-Oseen vortices with individual strength $\Gamma$  and separated by a distance $2 d_0$. In addition, ``mirror" vortices with strength $-\Gamma$ are placed symmetrically with respect to the $x$-axis, at a distance $2 L_0$ below the first pair, which causes the vortex system to move in the $x$-direction.  Due to viscosity,  the mirror vortices in this case do not represent the effect of a wall,  but they are added to create an open flow that corresponds to the viscous analogue of the flow system considered in the previous sections. In the following simulations, one million passive and collisionless particles are injected at random initial positions in a region covering the upper vortices. Then the particle and the fluid equations are solved for several turnover times, starting at $t=0$, until vortex merging occurs.  We use the same non-dimensionalization for length and time scales used in our study of the internal flow, except that here $d_0$ and  $\Omega_0$ are the {\it initial}  half-distance and angular velocity, respectively. The corresponding Reynolds number of the flow, Re$= \Omega_0 d_0^2/\nu$ (where $\nu$ is the kinematic viscosity), is  equal to $400$ and the Stokes number of the particles is $\St=0.07$. 

The flow domain is a two-dimensional periodic box, which allows us to use a Fourier series decomposition in both $x$ and $y$.   A second-order Adams-Bashforth algorithm is employed for the time integration of both the fluid and the particle equations, with a time-step calculated to satisfy the Courant-Friedrichs-Lewy condition \citep{Canuto1988}.   The fluid velocity at the particle position is interpolated by means of Shepard's method (inverse distance weighted interpolation). Using these techniques, we implemented two runs, corresponding to   
$\vepsil = 0.4$  and   $\vepsil = 0.2$, respectively. In the former case, the box size is equal to $15d_0$ in both the $x$ and $y$ directions, and  512$\times$512 Fourier modes are used. In the latter case, the size of the box was increased to $30 \, d_0$ in the $y$-direction to avoid spurious self-interactions due to the periodicity of the box for small $\vepsil$.

\begin{figure}
        \centering
                \includegraphics[width=0.8\textwidth]{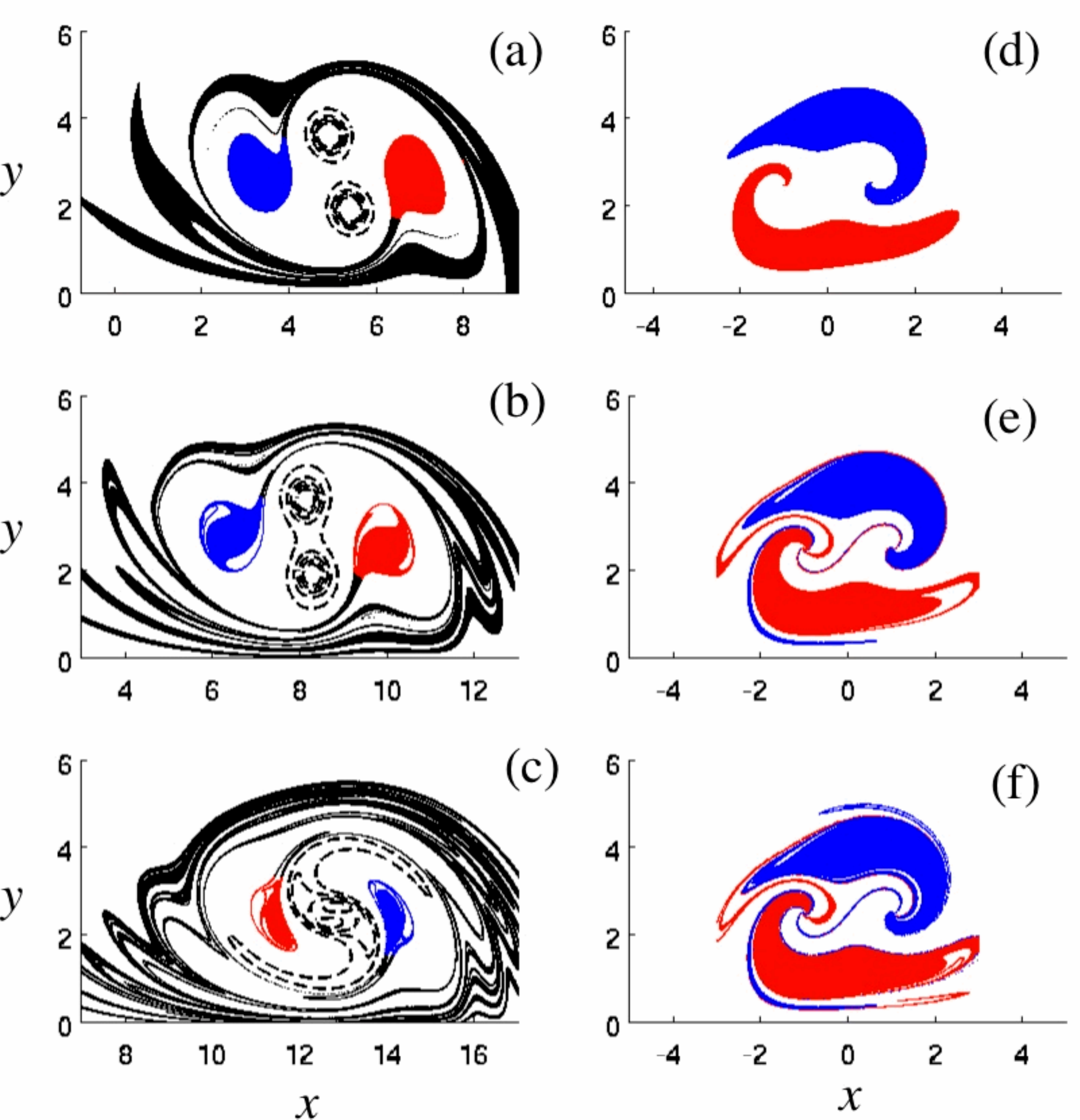}                 
\caption{\baselineskip 13pt
Fractal-like basin boundaries in Navier-Stokes simulations. (a-c) Starting with heavy particles uniformly distributed in a region that includes the vortices at  $t=0$, the panels show the position of the particles at times $t=7.1$ (a), $12.4$ (b) and $19.4$ (c). For the purpose of this illustration, particles marked with blue and red are considered trapped by the corresponding attractors, and dashed lines are iso-vorticity contours indicating the position of the vortices (same iso-values in all panels).  (d-f) Initial positions of the trapped particles of corresponding colors in panels (a), (b), and (c), respectively.  The symmetry line is located at $y=0$ and only the top two vortices are shown. The flow Reynolds number  Re is equal to $400$. The other parameters are $\St=0.07$ and $\vepsil = 0.4$. Note that the set defined by the initial conditions of the trapped particles becomes filamented as time increases. }
        \label{NS1}
\end{figure}

Figure \ref{NS1}(a) shows the particle cloud  for the $\vepsil = 0.4$ run  at time $t=7.1$.  Two clusters of particles are visible near the vortices (marked blue and red blobs). One can check that the particles follow the vortices until merging takes place and hence are temporarily trapped. The initial positions of the colored particles are indicated with the corresponding colors in Fig.\ \ref{NS1}(d). The overall shape of this basin is roughly comparable to that of the vortex pair of the inviscid fluid (Fig.\ \ref{BasinsSt0.07}(e)). This suggests that the clustering of particles seen in the Navier-Stokes simulation has the same dynamical origin as the trapping phenomenon studied in Secs.\ \ref{secAsym} and \ref{4vortex}.
The basin boundary of Fig.\ \ref{NS1}(d)  is smooth, however, since it corresponds to short simulation times. Figures \ref{NS1}(b) and \ref{NS1}(c) show particle clouds (at $t=12.4$ and $t=19.4$ respectively), which correspond to two typical structures  of the vorticity field (eight shape and spiral shape).  The initial positions of the trapped particles are indicated in Figs.\ \ref{NS1}(e) and \ref{NS1}(f). The basin boundaries now display a more filamentary structure, rather similar to the potential flow case (Fig.\ \ref{BasinsSt0.07}(e)). However, this structure cannot show very thin filaments, as viscosity causes vortex merging. Indeed, the spiral structure wrapped around the trapped particles (dashed lines in Fig.\ \ref{NS1}(c)) is temporary and is  eventually  smoothed out by viscous diffusion, which centrifugates the particles away.

\begin{figure}
        \centering
                \includegraphics[width=0.8\textwidth]{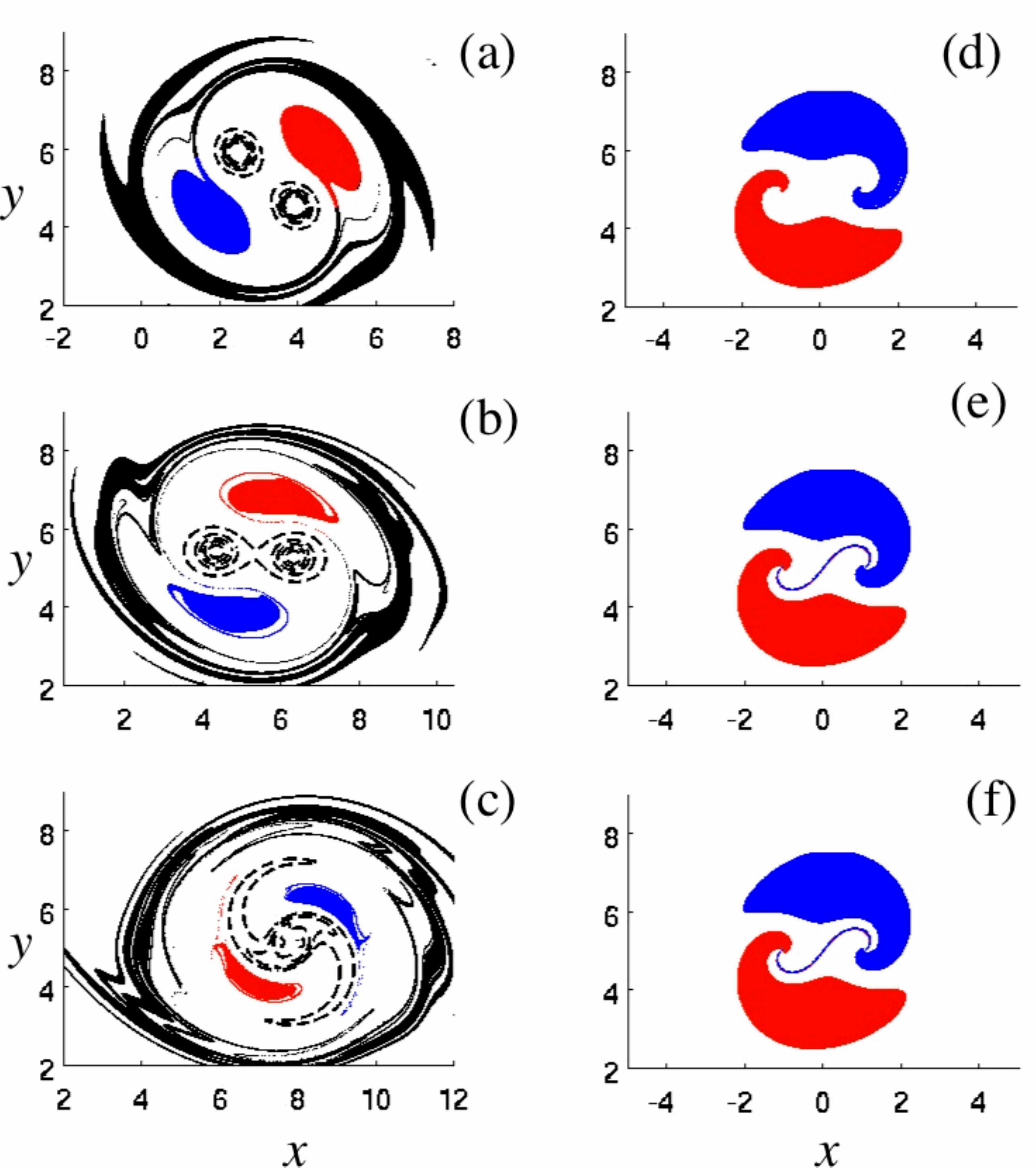} 
\caption{\baselineskip 13pt
Counterpart of Fig.\ \ref{NS1} for smooth basin boundaries.  Here, $\St$ and Re are the same as in Fig.\ \ref{NS1} and $\vepsil = 0.2$, which is above the critical distance to the symmetry line for the basin boundaries of the corresponding potential flow to become smooth. The particle distributions in panels (a-c) are represented at times $t=8.3$ (a), $14.9$ (b), and $22.0$ (c). The corresponding colored regions in panels (d-f) defined by the initial positions of the particles trapped remain non-filamented as time increases,  which further illustrates the agreement between potential flow predictions and Navier-Stokes simulations.
}
\label{NS4}
\end{figure}

The inviscid-fluid calculations of Sec.\ \ref{4vortex}  also suggest that the basin boundary should be smooth when  $\vepsil = 0.2$ (Fig.\ \ref{BasinsSt0.07}(d)). We have checked whether this could be observed also in the viscous case by setting $\vepsil = 0.2$ in our numerical calculations, corresponding to a distance $2/\vepsil = 10$ between the vortex pairs. Figure \ref{NS4} shows the particle cloud at three different times (left panels) corresponding to the three typical stages of vortex interaction, together with the initial positions of the trapped particles (right panels).  In this case the basins of attraction have  smooth boundaries,  as no filamentation is visible, and are therefore very similar to the portraits of Figs.\ \ref{BasinsSt0.07}(a) and \ref{BasinsSt0.07}(d).  This supports the conclusion that, to a good approximation, the potential flow theory correctly predicts the dynamics of heavy particles in this flow until vortex merging starts to occur.

\subsection{Effect of noise}

In the laboratory frame, the
dimensional form of the equation of motion for a heavy particle under thermal noise is \citep{Drossinos2005} 
\begin{equation}
\frac{d^2 \vec{X}_p}{dt^2} = \frac{1}{\tau_p}\left({\bf V}_f- \frac{d \vec{X}_p}{dt}\right) + {\bf f}(t), 
\label{noise} 
\end{equation}
where 
${\bf f}(t)$ is the random force per
unit of mass of the particle.
 The components of this force are assumed to 
 be of zero mean, Gaussian, and 
 delta-correlated in time:  
\begin{equation}
\left\langle f_i (t) f_j (t') \right\rangle = q\,\delta_{ij}\delta(t-t'), \quad   i,j\in\{x,y\}, 
\end{equation}
where 
 $\langle \cdot  \rangle$ denotes average and
$q$ is the strength of the 
force.  The fluctuation dissipation theorem  allows 
 relating
 $q$ to the diffusion coefficient ${\cal D}$ 
 as 
$q=2{\cal D}/\tau_{p}^2$
  \citep{Drossinos2005}.

Using the characteristic velocity magnitude 
and the characteristic length 
of the fluid flow, we can
write
 Eq.~(\ref{noise}) in dimensionless form:
\begin{equation}
\frac{d^2 \vec{X}_p}{dt^2}= \frac{1}{\St}\left({\bf V}_f- \frac{d \vec{X}_p}{dt}\right)  + \sqrt{2 \Delta} \, \xi (t),
\label{noise2}
\end{equation}
where 
$\Delta={1}/({\St^{2} \Pen })$ is the non-dimensional noise strength, 
$\Pen= \Omega_0 d_0^2/{\cal D}$
is the P\'eclet number, and $\xi(t)$ is a zero-mean normalized Gaussian white noise.

\begin{figure}
\centering
\includegraphics[width=0.65\textwidth]{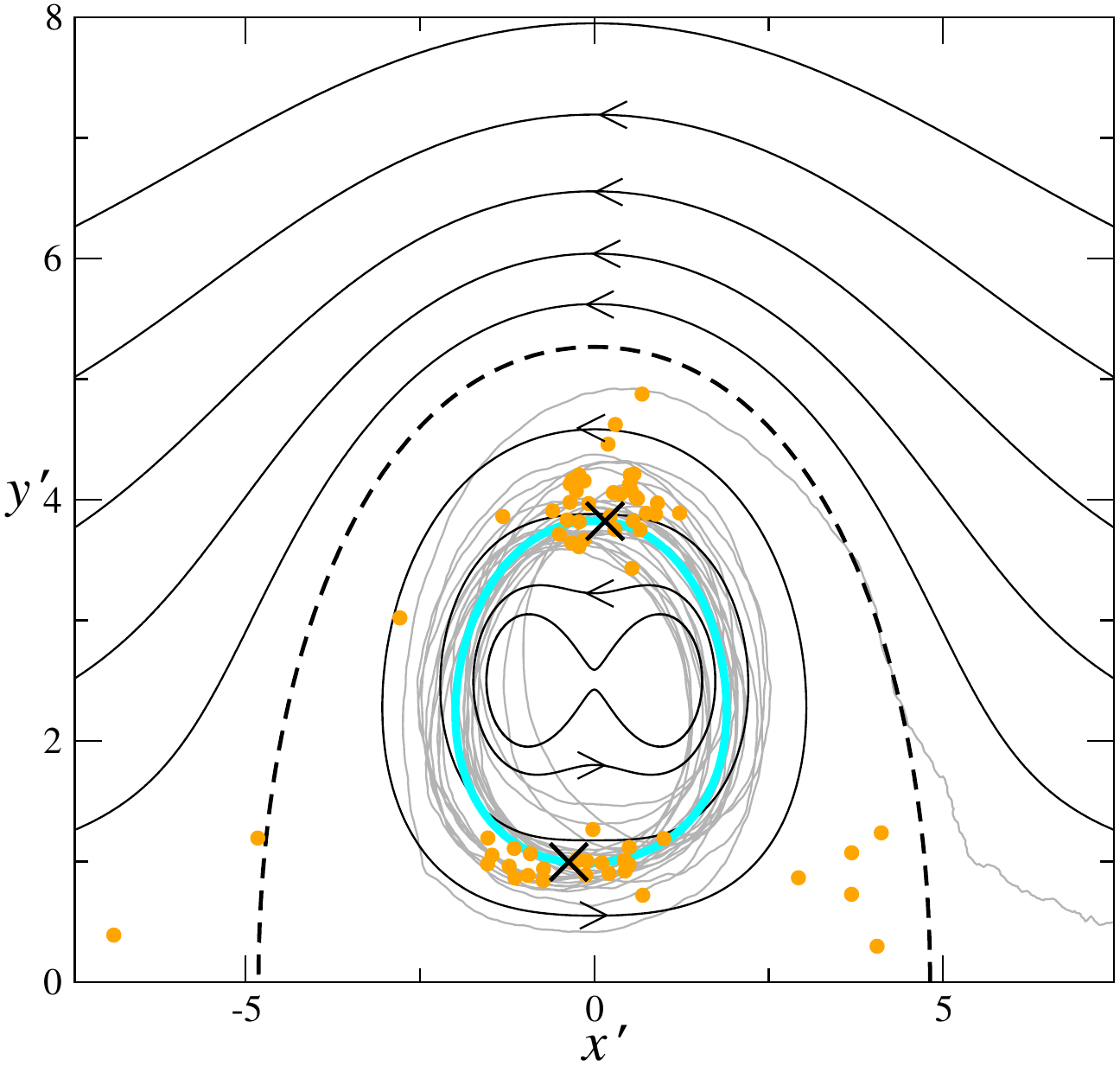}                                             
\caption{\baselineskip 13pt
Particle trapping in the presence of noise.
Heavy particles released in the open flow and subject to noise are shown after $30.25$ times the period of the fluid (orange dots) along with the projection of the full trajectory of one such particle (grey line). The attracting points of  the deterministic dynamics at the same instant are marked with cross symbols and their orbit is shown in cyan. Also shown are the instantaneous streamlines (continuous back lines) and external separatrix (dashed black line) of the deterministic flow dynamics.
The parameters are $\varepsilon=0.4$, $\St=0.006$, and (for the presence of noise) $\Pen=600$.}
\label{streamnoise} 
\end{figure}

\begin{figure}
\centering
\includegraphics[width=0.65\textwidth]{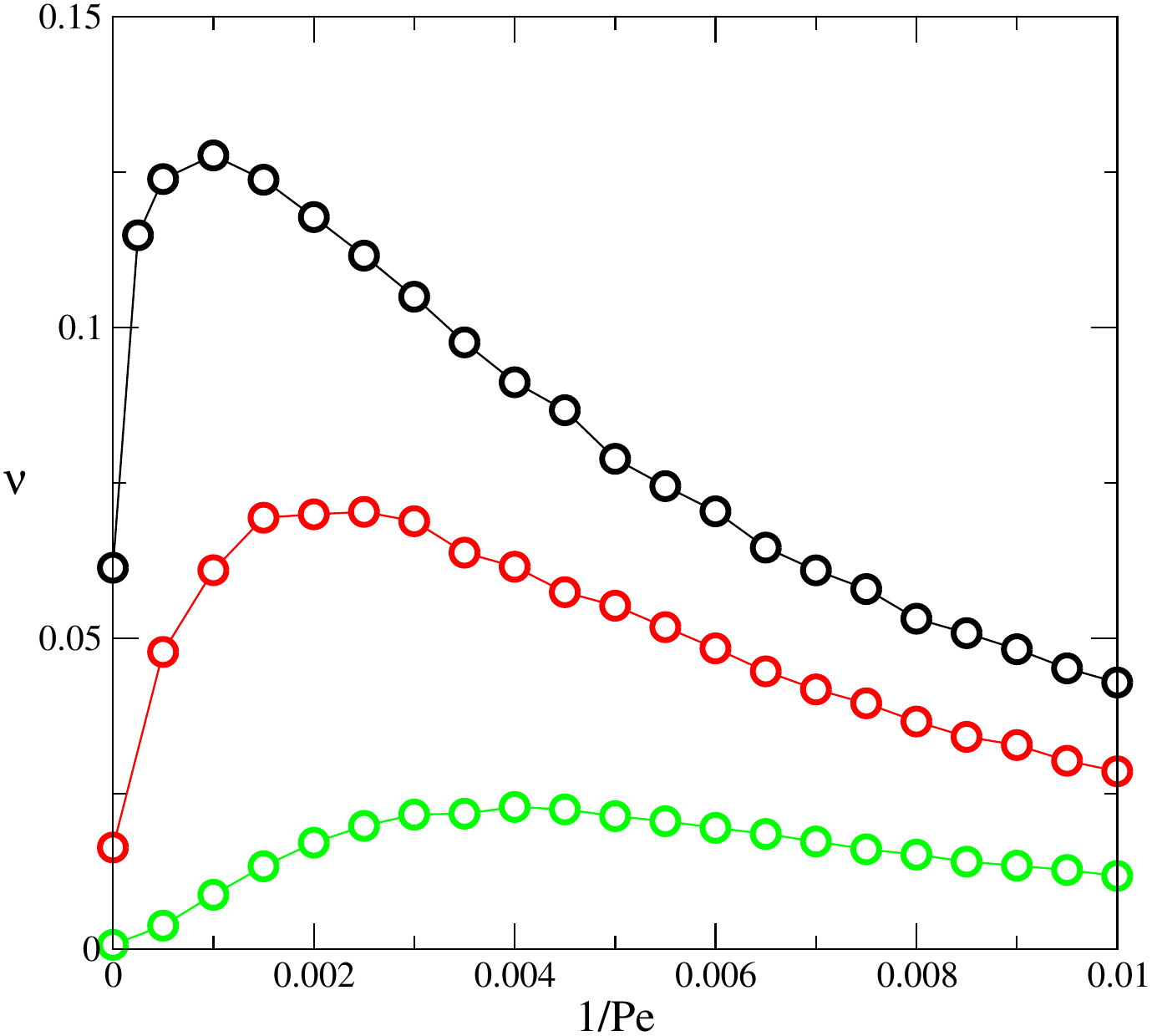} 
\caption{\baselineskip 13pt
Fraction $\nu$ of particles trapped as a function of the noise intensity $1/\Pen$. The different curves correspond to $\St=0.005$ (black), $0.007$ (red), and $0.010$ (green),  for $\varepsilon=0.4$.
Each data point corresponds to a total of $2\times 10^5$ trajectories for particles released from a uniform distribution along the line segment $x=5$ and $0\le y\le 0.5$.
As a criterion for trapping in the presence of noise, we regarded as trapped the particles that performed 
at least $40$ revolutions around the corresponding attracting center. }
\label{diagramnoise} 
\end{figure}

We have
explored
 numerically 
 the possibility of trapping 
 in the presence of
 noise. Figure \ref{streamnoise} shows the capture of 
 heavy particles
 released in the open component of the flow and whose motion is described by Eq.~(\ref{noise2}). A 
 systematic account of 
 the effects of noise for different Stokes numbers
 is summarized in Fig.~\ref{diagramnoise}, 
 where we show as a function of 1/Pe
 the fraction of particles released  in the open 
 flow  that are  trapped. 
The axis 1/Pe = 0 corresponds to the 
noiseless
case. 
Remarkably, as the noise intensity  
(i.e.,  1/Pe) 
increases from zero,  the percentage of particles trapped also increases;  this percentage only starts to decrease at sufficiently large noise intensities.
We therefore conclude 
that trapping is robust with respect to noise, and can 
in fact
be enhanced by noise. 
The enhancement  of particle trapping at intermediate noise levels may
be due to the fact that 
noise can 
cause
inertial particles
to cross the separatrix $\Sigma_3$ even 
when the inertia of the particles is  too large for this to occur  
in the
absence of noise.
 In contrast,  
 larger noise intensities 
 cause the particles 
 to move erratically
 and eventually inhibit trapping.


\section{Conclusions}
\label{concl}

The analytical calculations presented in this paper show that heavy particles released
 in the upstream flow of  
a  vortex pair (and its specular image,
modeled as a wall) 
 can be trapped by 
point attractors moving with the vortices.
The stability of these points is determined by
a balance between the centrifugal force (due to the rotation of vortices around each other) and the inward drag.  
It is observed that the dynamics of 
the
inertial particles can become transiently chaotic, as long as the distance between the vortex pairs (or, equivalently, the distance to the wall) is below a critical value that depends on the particle Stokes number. 
This 
chaotic behavior results in fractal basin boundaries for the attracting points,
which was verified 
for specific parameter choices
by showing that the
fractal dimension of the basin boundaries becomes larger than three in the four-dimensional phase space as soon as 
our analytical
criterion 
predicting a heteroclinic tangle near the separatrix $\Sigma_2$ 
 is fulfilled.

This metamorphosis of the
basin boundaries 
has a dramatic consequence for particle dynamics:  one can no longer easily predict which particles will  be 
captured
by a given attracting point and which particles will be
captured
by the other attracting point or, when the external separatrix can also be crossed, go to infinity. That is, due to transient chaos and the fractal basin boundaries that come with it, the particle dynamics exhibits final state sensitivity. Moreover, particles injected at different locations of the flow domain can undergo mixing in the the neighborhood of the chaotic set prior to converging to their final states (either of the attracting points or infinity). In other words, particles are mixed before being either trapped or centrifuged away.

For an observer translating with the vortices, the flow consists of a portion formed by closed streamlines and a portion  formed by open streamlines further away from the vortices, with separatrix $\Sigma_3$ at the boundary between them.  The trapping of heavy particles released in the open part of the flow requires that 
both
the separatrix $\Sigma_3$ and the separatrix $\Sigma_2$ 
of the flow 
be permeable with respect to the particles. 
We have shown that, 
 in order to become permeable to heavy particles,
 in the absence of gravity
   the separatrices have to not only break 
   but also give rise to heteroclinic tangles, 
   which occurs when the flow unsteadiness induced  by the wall is sufficiently strong. 
 Therefore, the wall has a double role: it not only causes the flow to be open, but it also allows particles to cross the separatrices and eventually be trapped in the neighborhood of the vortices.  We note that, while inertia is necessary for the formation of attractors, the larger the Stokes number the more difficult it is for the separatrices to be crossed (i.e., closer proximity to the wall is required).

The theory we established 
using perturbative velocity fields allowed us to generate a global trapping diagram 
(Fig.\ \ref{StEpsiXFIG}), which can be used to predict particle trapping in the 
$\St$--$\varepsilon$ plane for small particle Stokes number $\St$ and small inverse distance 
to the wall $\varepsilon$. Comparisons between this diagram and numerical simulations 
using the exact four-vortex system are excellent.  
This analysis reveals the mechanism underlying the trapping of aerosols from the open flow in this system. In the absence of gravity, it is the emergence of
heteroclinic tangles 
induced by the wall 
that
is responsible for trapping from the open flow.  
Further analysis 
demonstrated that,
in the presence of gravity, trapping  from the open flow is also possible 
without the formation of a heteroclinic tangle in the vicinity of the external separatrix.

 The robustness of particle trapping from the open flow
was verified 
by considering systematically the effects of
 gravity, noise, and viscosity. Trapping 
persists
 in the presence of
 gravity 
 for any orientation of the wall,
 provided that the settling velocity is not too large.
In particular, 
for
 a non-vertical wall the perturbative analysis could be readily generalized, and three 
kinds of
 behavior  
were
shown to exist 
(summarized in the
trapping 
diagram of Fig.\ \ref{diagramgravity} for a horizontal wall).  In this case gravity can cause the particles to cross the
external separatrix in a non-chaotic manner, and be trapped permanently by a limit cycle next to it. This behavior, which requires that
particles be allowed to slip on the wall (no deposition),  
is a form of non-chaotic trapping 
from the open flow. 
The limit cycle was observed to coexist with the attracting
points  near the vortices: particles released in the closed portion of the flow
can either spiral in toward one of the attracting
points or spiral out toward the limit cycle.

Numerical simulations of the Navier-Stokes equations 
showed that, when the fluid is viscous,  
the attracting points persist until vortex merging starts to occur.
When vortex coalescence takes place, the attracting points vanish and particles are centrifuged away, 
as expected.
But prior to the vortex merging, the overall structure of the ``attraction basins'' is rather similar to the  basins in the inviscid case,
provided that the flow Reynolds number is large enough.
Finally, 
 trapping 
also persists in the presence of noise.
Using the exact potential flow, we observed that 
  Brownian heavy 
  particles 
 can be trapped
  for at least several tens of periods of the background fluid flow,
  provided that the P\'eclet number is sufficiently large.

The particles in this study 
were taken to be sufficiently dilute so that their effect on the fluid could be neglected (``one-way coupling"). This assumption is not valid for large particle loadings, especially in zones where particles accumulate. Inertial particles have been shown to influence vortex pairing in mixing layers \citep{Meiburg2000,Wallner2002}. It could therefore be interesting to extend this study to investigate the effect of  the dispersed phase on the vortex pairing phenomenon considered here. In the same vein, particle collisions, which
were neglected in the present study, are known to influence the trapping process \citep{Medrano2008,Zahnow2009}. 
In addition, it would be interesting to consider particles with density comparable with the fluid density and investigate the effect of the Boussinesq-Basset, added-mass, and lift forces on the trapping process (see, for example, \cite{Daitche2011}, \cite{Sapsis2008}, \cite{DeLillo2008}, and \cite{Drotos2011}).
Such an extended analysis of inertial particle dynamics  
is  among the topics for future exploration
that we hope our results will  
encourage researchers to pursue.

\section*{Acknowledgement}

The authors thank I.~Fouxon for useful discussions and  E.~Hicks for providing feedback on the manuscript.
This research was supported by the National Science Foundation  through Grant No.\ PHY-1001198. 
 R.D.V.\ acknowledges financial support from CNPq (Brazil).

 $\phantom{.}$
\newpage 
 
$\phantom{.}$
\newpage 

$\phantom{.}$
\newpage 

$\phantom{.}$
\newpage 
 
$\phantom{.}$
\newpage 
 
$\phantom{.}$
   
$\phantom{.}$
\newpage

$\phantom{.}$
\newpage

$\phantom{.}$
\newpage

$\phantom{.}$
\newpage

$\phantom{.}$
\newpage
 
$\phantom{.}$
\newpage

$\phantom{.}$
\newpage

\begin{thebibliography}{99} 
\baselineskip 13.5pt


\bibitem[Angilella(2010)]{Angilella2010}
{\sc Angilella, J.-R.} 2010 Dust trapping in vortex pairs. {\em Physica D \/} {\bf 239}, 1789--1797.

\bibitem[Angilella(2011)]{Angilella2011}
{\sc Angilella, J.-R.} 2011 Asymptotic properties of wall-induced chaotic mixing in point vortex pairs. {\em Phys. Fluids\/} {\bf 23}, 113602.
 

\bibitem[Antonsen \& Ott(1991)]{Antonsen1991} 
{\sc  Antonsen, T.M. \& Ott, E. } 1991 Multifractal power spectra 
of passive scalars convected by chaotic fluid-flows. {\em Phys. Rev. A \/} {\bf 44}, 851--857.
         
\bibitem[Arnold (1965)]{Arnold1965}
{\sc Arnold, V. } 1965  Sur la topologie des \'ecoulements stationnaires des fluides parfaits. {\em C. R. Acad. Sci. Paris A \/} {\bf 261}, 17--20.

\bibitem[Babiano {\em et~al.\/}(2000)]{Babiano2000}
{\sc Babiano, A., Cartwright, J.~H.~E., Piro, O. \& Provenzale, A.}
  2000 Dynamics of a small neutrally buoyant sphere in a fluid and targeting in Hamiltonian systems. {\em Phys. Rev. Lett.\/} {\bf 84}, 5764--5767.

 
\bibitem[Balkovsky {\em et~al.\/}(2001)]{Balkovsky2001}
{\sc Balkovsky, E., Falkovich, G. \& Fouxon, A.} 2001 Intermittent distribution of
inertial particles in turbulent flows.   {\em Phys. Rev. Lett.   \/} {\bf 86}, 2790--2793.


\bibitem[Barge \& Sommeria(1995)]{Barge1995}
{\sc  Barge, P. \& Sommeria, J.} 1995  Did planet formation begin inside persistent gaseous vortices? {\em Astron. Astrophys.  \/} {\bf 295}, L1--L4.

 
\bibitem[Bec(2003)]{Bec2003}
{\sc Bec, J.} 2003 Fractal clustering of inertial particles in random flows.   {\em Phys.
Fluids \/} {\bf 15}, L81--L84.


\bibitem[Benczik(2002)]{Benczik2002}
{\sc Benczik, I.~J., Toroczkai, Z. \& T\'el, T.} 2002 Selective sensitivity of open chaotic flows on inertial tracer advection: Catching particles with a stick.   {\em Phys.
Rev. Lett. \/} {\bf 89}, 164501.


\bibitem[Benczik(2003)]{Benczik2003}
{\sc Benczik, I.~J., Toroczkai, Z. \& T\'el, T.} 2003 Advection of finite-size particles in open flows.   {\em Phys. Rev. E \/} {\bf 67}, 036303.

 
\bibitem[Canuto {\em et al.} (1988)]{Canuto1988}
{\sc Canuto, C., Hussaini, M.~Y., Quarteroni, A. \& Zang, T.~A.} 1988 {\em Spectral
  Methods in Fluid Dynamics}.  Springer-Verlag, Berlin.

\bibitem[Carton {\em et~al.\/}(2002)]{Carton2002}
{\sc Carton, X., Maze, G. \& Legras, B.} 2002 A two-dimensional vortex merger
  in an external strain field. {\em J. Turbul.\/} {\bf 3}, 045.


\bibitem[Cartwright {\em et al.}(2010)]{Cartwright2010}
{\sc Cartwright, J., Feudel, U.,  Karolyi, G., De Moura, A.,  Piro, O. \& Tel, T.} 2010 Dynamics of finite-size particles in chaotic
fluid flows. In {\em Nonlinear Dynamics and Chaos: Advances and Perspectives} (ed. {\sc M. Thiel}).  Springer-Verlag, Berlin.


\bibitem[Cerretelli \& Williamson(2003)]{Cerretelli2003}
{\sc Cerretelli, C. \& Williamson, C.~H.~K.} 2003 The physical mechanism for
  vortex merging. {\em J. Fluid Mech.\/} {\bf 475}, 41--77.
 

\bibitem[Cuzzi {\em et al.}(2001)]{Cuzzi2001}
{\sc Cuzzi, J.~N., Hogan, R.~C., Paque, J.~M. \& Dobrovolskis, A.~R.} 2001 Size-selective
concentration of chondrules and other small particles in protoplanetary nebula
turbulence. {\em Astrophys. J. \/} {\bf 546},  496--508.


\bibitem[Daitche \& T\'el(2011)]{Daitche2011}
  {\sc Daitche, A. \& T\'el, T.} 2011 Memory effects are relevant for chaotic advection of inertial particles. {\em Phys. Rev. Lett.\/} {\bf 107}, 244501.
 

\bibitem[De Lillo {\em et al.}(2008) ]{DeLillo2008}
{\sc De Lillo, F., Cecconi, F., Lacorata, G. \& Vulpiani, A.} 2008 Sedimentation speed of inertial particles in laminar and turbulent flows. {\em Europhys. Lett. \/} {\bf 84},  40005.


\bibitem[Drossinos \& Reeks(2005)]{Drossinos2005}
{\sc Drossinos, Y. \& Reeks, M.~W.} 2005 Brownian motion of finite-inertia particles in a simple shear flow. {\em Phys. Rev. E\/} {\bf 71}, 031113.


\bibitem[Drotos \& T\'el(2011)]{Drotos2011}
{\sc Drotos, G. \& T\'el, T.} 2011 Chaotic saddles in a gravitational field: The case of inertial particles in finite domains. {\em Phys. Rev. E\/} {\bf 83}, 056203.
 
 
\bibitem[Duncan {\em et al.}(2005)]{Duncan2005}
{\sc  Duncan, K., Mehlig, B., Ostlund, S. \&  Wilkinson, M.} 2005 Clustering by mixing flows. {\em  Phys. Rev. Lett.\/} {\bf 95}, 240602.

\bibitem[Falkovich {\em et al.}(2001)]{Falkovich2001}
{\sc  Falkovich, G., Gawedski, K. \& Vergassola, M. } 2001 Particles and fields in fluid turbulence. {\em  Rev. Mod. Phys. \/} {\bf 73}, 913--975.
 

\bibitem[Falkovich {\em et al.}(2002)]{Falkovich2002}
{\sc  Falkovich, G., Fouxon, A. \& Stepanov, M.~G. } 2002 Acceleration of rain initiation by
cloud turbulence. {\em  Nature \/} {\bf 419}, 151--154.
 

\bibitem[Fessler {\em et al.}(1994)]{Fessler1994}
{\sc  Fessler, J.~R., Kulick, J.~D. \& Eaton, J.~K.  } 1994 Preferential concentration of heavy
particles in a turbulent channel flow. {\em  Phys. Fluids \/} {\bf 6}, 3742--3749.


\bibitem[Fouxon(2012)]{Fouxon2012}
{\sc Fouxon, I.} 2012 Distribution of particles and bubbles in turbulence at a small Stokes number. {\em Phys. Rev. Lett.\/} {\bf 108}, 134502.

 
\bibitem[Gatignol(1983)]{Gatignol1983}
{\sc  Gatignol, R.} 1983 {The {F}ax\'en formulae for a rigid particle in an unsteady non-uniform {S}tokes flow.}
{\em J. M\'ec. Th\'eor. Appl.  \/} {\bf 1}, 143--160.
 
\bibitem[Gautero (1985)]{Gautero1985}
{\sc Gautero, J.L.} 1985  Chaos lagrangien pour une classe d'\'ecoulements de Beltrami. {\em C. R. Acad. Sci. S. II.  \/} {\bf 301}(15), 1095--1098.

\bibitem[Gelfreich(1997)]{Gelfreich1997}
{\sc Gelfreich,  V.~G.}  1997 {
Melnikov method and exponentially small splitting of separatrices.}
{\em Physica D \/} {\bf 101}, 227--248.
 

\bibitem[Grassberger(1986)]{Grassberger1986}
 {\sc Grassberger, P.}  1986 Estimating the fractal dimensions and entropies of strange attractors. In {\em Chaos} (ed. {\sc A.V. Holden}).
 Manchester University Press, Manchester.

  
\bibitem[Guckenheimer \& Holmes(1983)]{Guckenheimer1983}
{\sc Guckenheimer, J. \& Holmes, P.}  1983 {\em Nonlinear oscillations, dynamical systems, and bifurcations of vector fields}. Springer, New York.

\bibitem[Haller \& Sapsis(2008)]{Haller2008}
{\sc  Haller, G. \& Sapsis, T.}  2008 
{Where do inertial particles go in fluid flows?}
{\em Physica D \/} {\bf 237}, 573--583.


\bibitem[Haller \& Sapsis(2010)]{Haller2010}
{\sc  Haller, G. \& Sapsis, T.}  2010 
{Localized instability and attraction along invariant manifolds.}
{\em SIAM J. Applied Dynamical Systems \/} {\bf 9}(2), 611--633.

\bibitem[Ijzermans \& Hagmeijer(2006)]{IJzermans2006}
{\sc Ijzermans, R.~H.~A. \& Hagmeijer, R.} 2006 Accumulation of heavy particles
  in {N}-vortex flow on a disk. {\em Phys. Fluids\/} {\bf 18}, 063601.


\bibitem[Liu {\em et~al.\/}(2010)]{Liu2010}
{\sc Liu, S.-J., Wei, H.-H., Hwang, S.-H. \& Chang, H.-C.}
  2010 Dynamic particle trapping, release, and sorting by microvortices on a substrate. {\em Phys. Rev. E\/} {\bf 82}, 026308.

 
\bibitem[McLaughlin(1988)]{McLaughlin1988}
{\sc McLaughlin, J.~B.} 1988 Particle size effects on lagrangian turbulence.
  {\em Phys. Fluids\/} {\bf 31}, 2544--2553.
 
\bibitem[Maxey(1987a)]{Maxey1987pof}
{\sc Maxey, M.~R.} 1987 The motion of small spherical particles in a cellular
  flow field. {\em Phys. Fluids\/} {\bf 30}, 1915--1928.

  
\bibitem[Maxey(1987b)]{Maxey1987jfm}
{\sc Maxey, M.~R.} 1987 The gravitational settling of aerosol particles in
		  homogeneous turbulence and random flow fields. {\em  J. Fluid Mech. \/} {\bf 174}, 441--465.

\bibitem[Maxey \& Corrsin(1986)]{Maxey1986}
{\sc Maxey, M.~R. \& Corrsin, S.} 1986 Gravitational settling of aerosol
  particles in randomly oriented cellular flow fields. {\em J. Atmos. Sci.\/}
  {\bf 43}, 1112--1134.


\bibitem[Maxey \& Riley(1983)]{Maxey1983}
{\sc Maxey, M.~R. \& Riley, J.~J.} 1983 Equation of motion for a small rigid
  sphere in a non uniform flow. {\em Phys. Fluids\/} {\bf 26}, 883--889.


\bibitem[Maze {\em et~al.\/}(2004)]{Maze2004}
{\sc Maze, G., Carton, X. \& Lapeyre, G.} 2004 Dynamics of a 2{D} vortex
  doublet under external deformation. {\em Regul. Chaotic Dyn.\/} {\bf
  9}, 477--497.

\bibitem[Medrano {\em et~al.\/}(2008)]{Medrano2008}
{\sc Medrano, R.~O., Moura, A., Tel, T., Caldas, I.~L. \& Grebogi, C.} 2008
  Finite-size particles, advection, and chaos: A collective phenomenon of
  intermittent bursting. {\em Phys. Rev. E\/} {\bf 78}, 056206.


\bibitem[Mehlig {\em et~al.\/}(2005)]{Mehlig2005}
{\sc Mehlig, B., Wilkinson, M., Duncan, K., Weber, T. \& Ljunggren, M.} 2005
  Aggregation of inertial particles in random flows. {\em Phys. Rev. E\/} {\bf 72}, 051104.

\bibitem[Meiburg {\em et~al.\/}(2000)Meiburg, Wallner, Pagella, Riaz, Hartel \&
  Necker]{Meiburg2000}
{\sc Meiburg, E., Wallner, E., Pagella, A., Riaz, A., Hartel, C. \& Necker, F.}
  2000 Vorticity dynamics of dilute two-way-coupled particle-laden mixing
  layers. {\em J. Fluid Mech.\/} {\bf 421}, 185--227.

 
\bibitem[Meyer \& Deglon(2011)]{Meyera2011}
{\sc  Meyer, C.~J. \&   Deglon, D.~A.  } 2011 Particle collision modeling - A review. {\em Miner. Eng.\/} {\bf 24}, 719--730.


\bibitem[Michaelides(1997)]{Michaelides1997}
{\sc Michaelides, E.~E.} 1997 The transient equation of motion for particles, bubbles, and droplets. {\em J Fluid Eng.-T. Asme \/} {\bf 119}, 233-247.

\bibitem[Olla(2010)]{Olla2010}
{\sc Olla, P.} 2010 Preferential concentration versus clustering in inertial particle transport
by random velocity fields. {\em Phys. Rev. E \/} {\bf 81}, 016305.
 

\bibitem[Ottino(1989)]{Ottino1989}
{\sc Ottino, J.}  1989 {\em The kinematics of mixing: stretching, chaos and transport.} Cambridge University Press, Cambridge, UK.

\bibitem[Pasquero {\em et~al.\/}(2003)]{Pasquero2003}
{\sc Pasquero, C., Provenzale, A. \&  Spiegel, E. A.} 2003  
Suspension and fall of heavy particles in random two-dimensional flow.
{\em Phys. Rev. Lett.} {\bf 91}, 054502.

\bibitem[Pushkin {\em et~al.\/}(2011)]{Pushkin2011}
{\sc Pushkin, D.~O., Melnikov, D.~E. \&  Shevtsova, V.~M.} 2011 Ordering of Small Particles in One-Dimensional Coherent Structures by Time-Periodic Flows. {\em Phys. Rev. Lett.\/} {\bf 106}, 234501.

\bibitem[Rom-Kedar {\em et~al.\/} (1990)]{RomKedar1990}
{\sc Rom-Kedar, V., Leonard, A. \& Wiggins, S.} 1990
  An analytical study of transport, mixing and chaos in an unsteady
  vortical flow.  {\em J. Fluid Mech.\/} {\bf 214}, 347--394.

\bibitem[Rubin {\em et~al.\/}(1995)]{Rubin1995}
{\sc Rubin, J., Jones, C.~K.~R.~T. \& Maxey, M.} 1995 Settling and asymptotic
  motion of aerosol particles in a cellular flow field. {\em J. Nonlinear
  Sci.\/} {\bf 5}, 337--358.

\bibitem[Sapsis \& Haller(2008)]{Sapsis2008}
  {\sc Sapsis, T. \& Haller, G.} 2008 Instabilities in the dynamics of neutrally buoyant particles. {\em Phys. Fluids\/} {\bf 20}, 017102.


\bibitem[Sapsis \& Haller(2009)]{Sapsis2009}
  {\sc Sapsis, T. \& Haller, G.} 2009 Inertial particle dynamics in a hurricane. {\em J. Atmosph. Sci.\/} {\bf 66}, 2481--2492.


\bibitem[Sapsis \& Haller(2010)]{Sapsis2010}
{\sc Sapsis, T. \& Haller, G.} 2010 Clustering criterion for inertial particles in two-dimensional time-periodic and three-dimensional steady flows. {\em Chaos\/} {\bf 20}, 017515.


\bibitem[Shaw(2003)]{Shaw2003}
{\sc Shaw, R.~A.} 2003 Particle-turbulence interactions in atmospheric clouds. {\em Annu. Rev. Fluid Mech.\/} {\bf 35}, 183--227.


\bibitem[Squires \& Eaton(1991)]{Squire02}
{\sc Squires, K.~D. \& Eaton, J.~K.} 1991 Preferential concentration of
  particles by turbulence. {\em Phys. Fluids\/} {\bf 3}, 1169--1178.


\bibitem[Stommel(1949)]{Stommel1949}
{\sc Stommel, H.} 1949 Trajectories of small bodies sinking slowly through convection cells.
 {\em J. Marine Res.\/} {\bf 8}, 24--29.

\bibitem[Vilela \& Motter(2007)]{Vilela2007}
{\sc Vilela, R.~D. \& Motter, A.~E.} 2007 Can aerosols be trapped in open flows? {\em Phys. Rev. Let.\/} {\bf 99}, 264101.


\bibitem[Wallner \& Meiburg(2002)]{Wallner2002}
{\sc Wallner, E. \& Meiburg, E.} 2002 Vortex pairing in two-way coupled,
  particle laden mixing layers. {\em Int. J. Multiphas. Flow\/} {\bf 28},
  325--346.


\bibitem[Wilkinson {\em et~al.\/}(2007)]{Wilkinson2007}
{\sc Wilkinson, M., Mehlig, B., Ostlund, S. \& Duncan, K.~P.} 2007 Unmixing in random flows. {\em Phys. Fluids\/} {\bf 19}, 113303.


\bibitem[Wilkinson {\em et~al.\/}(2010)]{Wilkinson2010}
{\sc Wilkinson, M., Mehlig, B. \& Gustavsson, K.} 2010 Correlation dimension of inertial particles in random flows. {\em Europhys. Lett.\/} {\bf 89}, 50002.


\bibitem[Zahnow \& Feudel(2009)]{Zahnow2009}
{\sc Zahnow, J.~C. \& Feudel, U.} 2009 What determines size distributions of heavy drops in a synthetic turbulent flow? {\em Nonlinear Proc. Geoph.\/} {\bf 16}, 677--690. 


\end{thebibliography}
\end{document}